\documentclass[12pt]{article}
\usepackage{psfig}
\begin{document}
\title{Near-threshold production of $a_0(980)$-mesons in $\pi N$ and
$NN$ collisions and $a_0/f_0$-mixing}
\author{
L.A. Kondratyuk,
E.L. Bratkovskaya\footnote{Institut f\"ur Theoretische Physik,
        Universit\"at Frankfurt, D-60054 Frankfurt, Germany},
V. Yu. Grishina\footnote{Institute for Nuclear Research,
        60th October Anniversary Prospect 7A, 117312 Moscow, Russia},\\
M. B\"uscher\footnote{Institut f\"ur Kernphysik,
        Forschungszentrum J\"ulich, D-52425 J\"ulich, Germany},
W.~Cassing\footnote{Institut f\"ur Theoretische Physik,
        Universit\"at Giessen, D-35392 Giessen, Germany},
 \ and H.~Str\"oher$^\ddag$ \\[5mm]
 Institute of Theoretical and Experimental Physics, \\
 B. Cheremushkinskaya 25, 117259 Moscow, Russia  }
\date{  }
\maketitle

\begin{abstract}
We consider near-threshold $a_0(980)$-meson production in $\pi N$ and
$NN$ collisions. An effective Lagrangian approach with one-pion
exchange is applied to analyze different contributions to the cross
section for different isospin channels. The Reggeon exchange mechanism
is also evaluated for comparison.  The results from $\pi N$ reactions
are used to calculate the contribution of the $a_0$ meson to the cross
sections and invariant $K \bar K$ mass distributions of the reactions
$pp\to pn K^+\bar K^0$ and $pp\to pp K^+K^-$. It is found that the
experimental observation of $a_0^+$ mesons in the reaction $pp\to pn
K^+\bar K^0$ is much more promising than the observation of $a_0^0$
mesons in the reaction $pp\to pp K^+K^-$.  Effects of isospin violation
in the reactions $pN \rightarrow d a_0$, $pd \rightarrow
\mathrm{^3He/^3H}\, a_0$, and $ dd \rightarrow \mathrm{^4He}\, a_0$,
which are induced by $a_0(980)$--$f_0(980)$ mixing, are also analyzed.
\end{abstract}


\newpage
\section{Introduction}

The structure of the lightest scalar mesons $a_0(980)$ and $f_0(980)$
is still under discussion (see, e.g.,
\cite{Clo}--\cite{Hadron99b} and references
therein). Different authors interpreted them as unitarized $q\bar{q}$
states,  as four-quark cryptoexotic states,  as $K\bar{K}$ molecules
or even as vacuum scalars (Gribov's minions).  Although it has been
possible to describe them as ordinary $q\bar{q}$ states (see
\cite{Montanet}--\cite{Narison}), other options cannot be ruled out
up to now.  Another problem is the possible strong mixing between the
uncharged $a_0(980)$ and the $f_0(980)$ due to a common coupling to
$K\bar K$ intermediate states
\cite{Achasov}--\cite{Grishina2001}.
This effect can influence the structure of the uncharged component of
the $a_0(980)$ and implies that it is important to perform a
comparative study of $a_0^0$ and $a_0^+$ (or $a_0^-$).  There is no
doubt that new data on $a_0^0$ and $a_0^+/a_0^-$ production in $\pi N$
and $NN$ reactions are quite important to shed new light on the $a_0$
structure and the dynamics of its production.

In our recent paper \cite{Grishina3} we have considered $a_0$ production
in the reaction $\pi N \to a_0N$ near the threshold and at GeV
energies. An effective Lagrangian approach as well as the Regge pole
model were applied to investigate different contributions to the
cross section of the reaction  $\pi N \to a_0N$.  In \cite{Brat01}
we have employed the latter results for an analysis of $a_0$ production
in $NN$ collisions. Furthermore, in \cite{Grishina2001} we have
considered the $a_0$--$f_0$ mixing in reactions involving the lightest
nuclei $d,{}^3$H, ${}^3$He, and ${}^4$He. Here we give an overview of
those results and present a comparative analysis of $a_0(980)$
resonance production and nonresonant background channels in the
reactions $\pi N \to a_0 N \to K \bar{K} N$ and $N N \to a_0NN \to K
\bar K NN$. Our study is particularly relevant to the current
experimental program at COSY (J\"ulich) \cite{COSY1}--\cite{COSY3}.

Our paper is organized as follows. In Section 2 we discuss the $K\bar K$
and $\pi \eta$ decay channels of the $a_0(980)$. An analysis of
$a_0(980)$ resonance production and nonresonant background in the
reactions $\pi N \to K \bar{K} N$ and $N N \to a_0NN \to K \bar K NN$
is presented in Section 3.  Section 4 is devoted to the calculation of the
cross sections for the reactions $NN \to NN a_0$ and $N N \to a_0NN \to
K \bar K NN$ in comparison to nonresonant $K \bar K$ production.  In
Section 5 we consider $a_0(980)$--$f_0(980)$ mixing and isospin violation
in the reactions $pN \rightarrow d a_0$, $pd \rightarrow
\mathrm{^3He/^3H}\, a_0$ and $ dd \rightarrow \mathrm{^4He}\, a_0$.

\section{The $K\bar K$ and $\pi \eta$ Decay Channels
of the $a_0(980)$}

   The $a_0(980)$ invariant mass distribution in $K\bar K$ and $\pi\eta$
modes can be parametrized by the well-known Flatt\'e formula
\cite{Flatte} which follows from analyticity and unitarity for the
two-channel $T$-matrix.

For example, in the case of the reaction $N N \to a_0NN \to K \bar K
NN$ the mass distribution of the final $K\bar K$ system can be written
as a product of the total cross section for $a_0$ production (with the
``running'' mass $M$) in the $NN\to NN a_0$ reaction and
the Flatt\'e mass distribution function
\begin{eqnarray}
\frac{d\sigma _{K\bar K}}{d M^2} (s,M) = \sigma_{a_0}(s,M)
\ C_F \frac{M_R \Gamma_{a_0 K\bar K}(M)}
{(M^2-M_R^2)^2
+ M_R^2 \Gamma_{\rm tot}^2(M)} \label{dsdmKK}
\end{eqnarray}
with the total width $\Gamma_{\rm tot}(M)=\Gamma_{a_0 K\bar K}(M)+
\Gamma_{a_0 \pi\eta}(M)$.
The partial widths
\begin{eqnarray}
&&\Gamma_{a_0 K\bar K}(M) = g_{a_0 K\bar K}^2 {q_{K\bar K}\over 8\pi M^2},
\nonumber\\
&&\Gamma_{a_0 \pi\eta}(M) = g_{a_0 \pi\eta}^2 {q_{\pi\eta}\over 8\pi M^2}
\label{width}\end{eqnarray}
are proportional to the decay momenta in the c.m. system
(in case of scalar mesons),
\begin{eqnarray}
&& q_{K\bar K} = {\left[(M^2-(m_{K}+m_{\bar K})^2)
(M^2-(m_{K}-m_{\bar K})^2)\right]^{1/2} \over 2M}, \nonumber\\
&& q_{\pi\eta}={\left[(M^2-(m_{\pi}+m_{\eta})^2)
(M^2-(m_{\pi}-m_{\eta})^2)\right]^{1/2} \over 2M}
\nonumber\end{eqnarray}
for a meson of mass $M$ decaying to  $K\bar K$ and $\pi\eta$,
correspondingly.  The branching ratios ${\rm Br}(a_0\to K\bar K)$ and
${\rm Br}(a_0\to \pi\eta)$ are given by the integrals of the Flatt\'e
distibution over the invariant mass squared $dM^2 = 2 M dM$:
\begin{eqnarray}
\phantom{a}\hspace*{-3mm}
&&{\rm Br}(a_0\!\to\! K\bar K)=\!\!\!\!\!\!\!\int\limits_{m_K
+m_{\bar K}}^{\infty}
\!\!\!\!\!\!\frac{dM \ 2\ M\ C_F \ M_R\ \Gamma_{a_0 K\bar K}(M)}
{(M^2-M_R^2)^2+M_R^2 \Gamma_{\rm tot}^2(M)},
\label{BrKK}  \\
\phantom{a}\hspace*{-3mm}
&&{\rm Br}(a_0\!\to\! \pi\eta)=\!\!\!\!\!\!\!\int\limits_{m_K
+m_{\bar K}}^{\infty}
\!\!\!\!\!\!\frac{dM \ 2 \ M \ C_F \ M_R \ \Gamma_{a_0 \pi\eta}(M)}
{(M^2-M_R^2)^2+M_R^2 \Gamma_{\rm tot}^2(M)} + \label{Brpieta}\\
\phantom{a}\hspace*{-3mm}
&&\!\!+\int\limits_{m_{\pi}+m_{\eta}}^{m_K+m_{\bar K}}
\!\!\frac{dM \ 2 \ M \ C_F \ M_R \  \Gamma_{a_0 \pi\eta}(M)}
{(M^2-M_R^2-M_R \Gamma_{a_0 K\bar K}(M))^2+M_R^2 \Gamma_{a_0 \pi\eta}^2(M)}.
\nonumber
\end{eqnarray}
The parameters $C_F, g_{K\bar K}, g_{\pi\eta}$ have to be fixed under the
constraint of the unitarity condition
\begin{eqnarray}
{\rm Br}(a_0 \to K\bar K) + {\rm Br}(a_0 \to \pi \eta)=1 \ .
\label{unitar}
\end{eqnarray}
Choosing the parameter $\Gamma_0=\Gamma_{a_0 \pi \eta}(M_R)$ in the
interval $50 - 100$ MeV (as given by the PDG \cite{PDG}),
one can fix the coupling $g_{\pi\eta}$ according to (\ref{width}).
In \cite{CrysBar98} a ratio of branching ratios has been reported,
\begin{eqnarray}
r(a_0(980))=\frac{{\rm Br}(a_0\to K\bar K)}{{\rm Br}(a_0\to
\pi\eta)}=0.23\pm 0.05,
\label{ratBr}
\end{eqnarray}
for $m_{a_0}=0.999$~GeV, which gives ${\rm Br}(a_0\to K\bar K)=0.187$.
In another recent study \cite{WA102} the WA102 collaboration
reported the branching ratio
\begin{equation}
\Gamma(a_0\to K\bar K) / \Gamma(a_0\to \pi \eta) = 0.166\pm
0.01\pm 0.02 , \label{eq02}
\end{equation}
which was determined from the measured branching ratio for the
$f_1(1285)$-meson.
In our present analysis we use the results from \cite{CrysBar98},
however, keeping in mind that this branching ratio ${\rm Br}(a_0\to K\bar K)$
more likely gives an ``upper limit'' for the $a_0\to K \bar K$ decay.

Thus, the two other parameters in the Flatt\'e distribution $C_F$ and
$g_{a_0 K\bar K}$ can be found by solving the system of integral
equations, for example, Eq. (\ref{BrKK}) for ${\rm Br}(a_0 \to K \bar K)
=0.187$ and the unitarity condition (\ref{unitar}).
For our calculations we choose either $\Gamma_{a_0 \pi \eta}(M_R)=70$~MeV
or 50 MeV, which gives two sets of independent parameters
$C_F, g_{a_0 K \bar K}, g_{a_0 \pi \eta}$ for a fixed branching
ratio ${\rm Br}(a_0 \to K \bar K)=0.187$:
\begin{eqnarray}
&&\hspace*{-8mm}
{\rm set} \ 1 \ \ (\Gamma_{a_0\pi\eta}=70~{\rm MeV}): \label{set1}\\
&&\phantom{a}\hfill  g_{a_0 K \bar K}=2.3 ~{\rm GeV}, \ g_{a_0 \pi \eta}=2.2~ {\rm GeV},
\ C_F=0.365 \nonumber\\
&&\hspace*{-8mm}
{\rm set} \ 2 \ \ (\Gamma_{a_0\pi\eta}=50~{\rm MeV}): \label{set2}\\
&&\phantom{a}\hfill  g_{a_0 K \bar K}=1.9~{\rm GeV}, \ g_{a_0 \pi \eta}=1.9~ {\rm GeV},
\ C_F=0.354.\nonumber
\end{eqnarray}
Note, that for the $K^+K^-$ or $K^0 \bar K^0$  final state one has
to take into account an isospin factor for the coupling constant, i.e.,
$g_{a_0 K^+K^-}=g_{a_0 K^0 \bar K^0} = g_{a_0 K\bar K}/\sqrt{2}$,
whereas $g_{a_0 K^+\bar K^0}=g_{a_0 K^- \bar K^0} = g_{a_0 K\bar K}$.

\section{The Reactions $\pi N\to a_0N$ and $\pi N\to K \bar K N$ }
\label{sec:pi-N}

\subsection{An effective Lagrangian Approach}

\begin{figure}[t]
\centerline{\psfig{figure=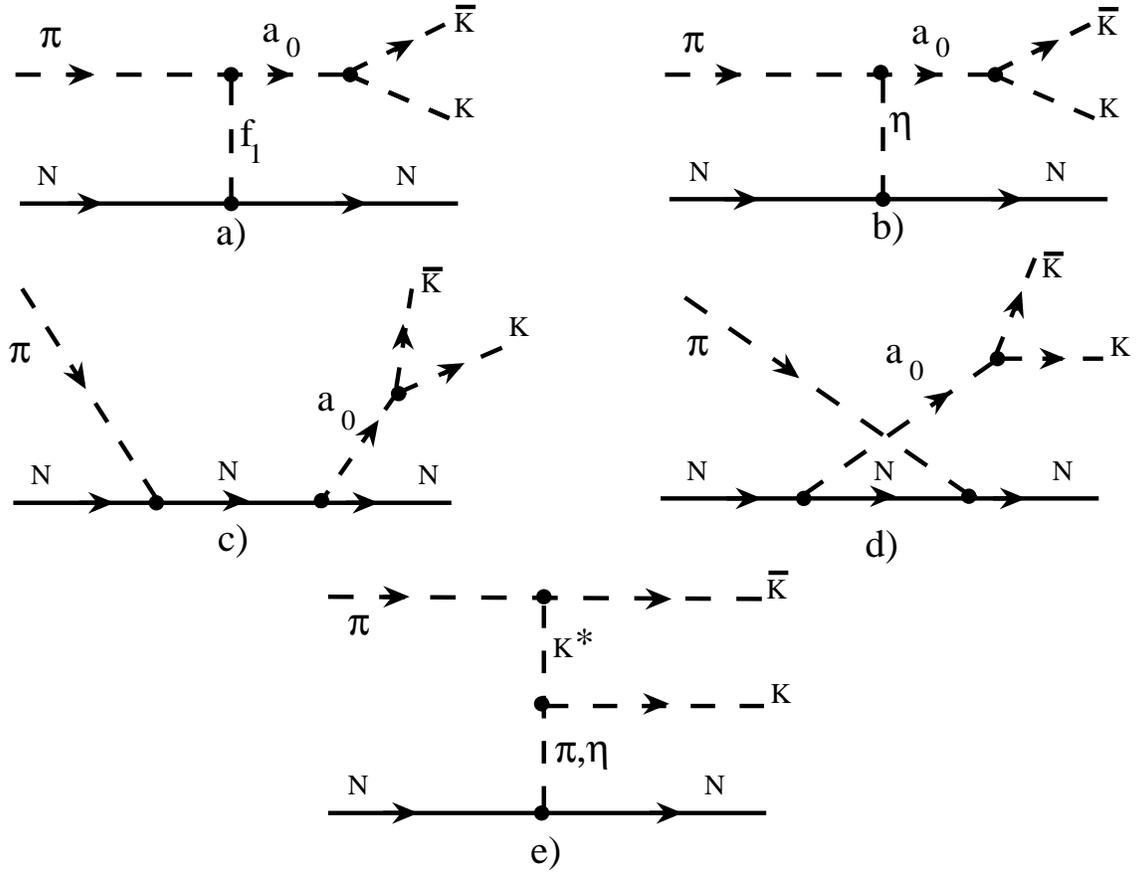,width=15cm}}
\vspace*{1cm}\caption{
The diagrams {\it a)-d)} for $a_0$ production in the reaction
$\pi N\rightarrow a_0 N \rightarrow \bar{K} K$ near threshold and a
diagram {\it e)} for nonresonant $\bar K K$ ``background'' production.
}\label{Fig1}
\end{figure}

   The most simple mechanisms for $a_0$ production in the reaction $\pi
N\rightarrow a_0 N$ near threshold  are described by the pole diagrams
shown in Fig. \ref{Fig1} {\it a} -- \ref{Fig1} {\it d}.  It is known
experimentally that the $a_0$ couples strongly to the channels
$\pi\eta$ and $\pi f_1(1285)$ because $\pi\eta$ is the dominant decay
channel of the $a_0$ while $\pi a_0$ is one of the most important decay
channels of the $f_1(1285)$ (\cite{PDG}). The amplitudes, which
correspond to the $t$-channel exchange of $\eta(550)$- and
$f_1(1285)$-mesons (see Fig. \ref{Fig1} {\it a} and Fig. \ref{Fig1} {\it b}), can be written as
\begin{eqnarray}
&&\hspace*{-4mm}M_\eta^t(\pi^-p\rightarrow a_0^- p) = g_{\eta\pi a_0}
g_{\eta NN}\ \bar u(p_2^\prime) \gamma_5 u(p_2) \times \nonumber\\
&\times& {1\over t-m_\eta ^2} \ F_{\eta\pi a_0}(t) F_{\eta NN}(t),
\label{eq2}\end{eqnarray}
\begin{eqnarray}
&&\hspace*{-4mm}M_{f_1}^t(\pi^- p\rightarrow a_0^- p) = g_{f_1\pi a_0}
g_{f_1NN}  \times \nonumber\\
&\times& (p_1+p_1^\prime)_\mu \ \left(g_{\mu\nu}-{q_\mu q_\nu\over
m_{f_1}^2}\right) \ \bar u(p_2^\prime) \gamma_\nu \gamma_5 u(p_2) \times
        \nonumber\\
&\times& {1\over t-m_{f_1}^2}\ F_{f_1\pi a_0}(t) F_{f_1NN}(t).
\label{eq3}
\end{eqnarray}
Here $p_1$ and $p_1^\prime$ are the four momenta of $\pi^-, a_0^-$,
whereas $p_2$ and $p_2^\prime$ are the four momenta of the initial and
final protons, respectively; furthermore, $q=p_2^\prime-p_2$,
$t=(p_2^\prime-p_2)^2$. The functions $F_j$ present form factors at the
different vertices $j$ ($j=f_1NN,\eta NN$), which are taken of the
monopole form
\begin{eqnarray}
F_j(t)=\frac{\Lambda_j^2-m_j^2}{\Lambda _j^2-t},
\label{form}
\end{eqnarray}
where $\Lambda_j$ is a cut-off parameter. In the case of $\eta$
exchange we use $g_{\eta NN}=6.1$, $\Lambda_{\eta NN}$=1.5 GeV from
\cite{Holinde} and $g_{\eta\pi a_0}$
is defined by (8). The contribution of
the $f_1$ exchange is calculated for two parameter sets; set $A$:
$g_{f_1 NN}=11.2$, $\Lambda_{f_1 NN}=1.5$~GeV from
\cite{Bonnf1}, set $B$: $g_{f_1 NN}=14.6$, $\Lambda_{f_1
NN}=2.0$~GeV from \cite{Kirchbach} and $g_{f_1a_0\pi}$=2.5
for both cases. The latter value for $g_{f_1 a_0 \pi}$ corresponds
to $\Gamma(f_1\to a_0\pi)=24$~MeV and ${\rm Br}(f_1\to a_0\pi)=34\%$.

\begin{figure}[t]
\psfig{figure=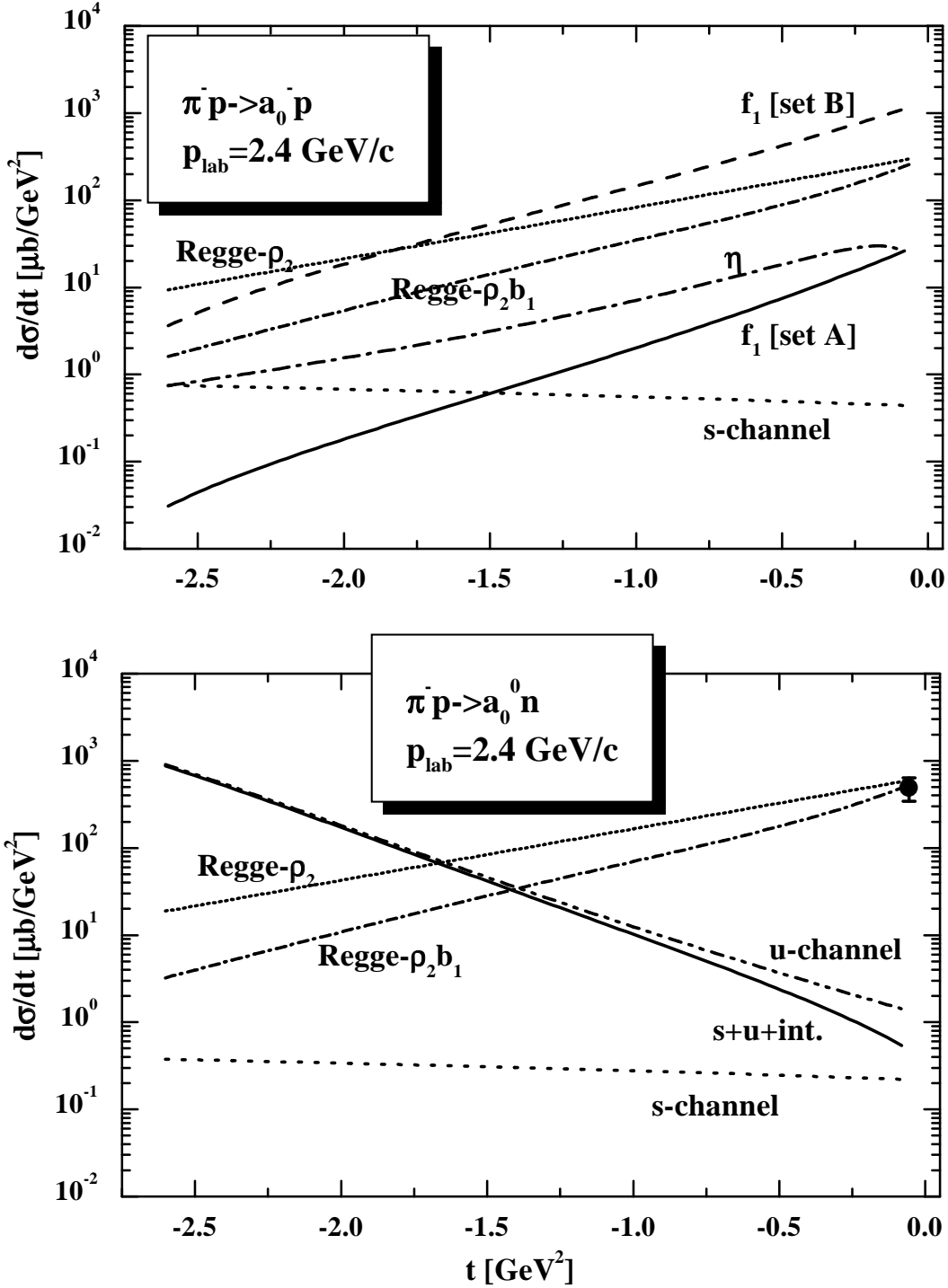,width=14cm}
\caption{
The differential cross sections $d\sigma/dt$ for the reactions
$\pi^-p\rightarrow  a_0^-p$ (upper part) and $\pi^-p\rightarrow a_0^0n$
(lower part) at 2.4 GeV$/c$.  The dash-dotted line corresponds to the
$\eta$ exchange, solid and dashed lines (upper part) show  the $f_1$
contributions within sets $A$ and $B$, respectively. The dotted and
dash-double-dotted lines indicate the $s$ and $u$ channels while the
solid line (lower part) describes the coherent sum of $s$- and $u$-
channel contributions. The short dotted and short dash-dotted lines
present the results within the $\rho_2$ and ($\rho_2, \ b_1$) Regge
exchange model, respectively (see text).
}\label{dsdt_pip}
\end{figure}

In Fig. \ref{dsdt_pip} (upper part) we show the differential cross
sections $d\sigma/dt$ for the reaction $\pi^-p\to a_0^- p$ at 2.4
GeV$/c$ corresponding to $\eta$ (long-dash-dotted) and $f_1$ exchanges
with set $A$ (solid line) and set $B$ (long-dashed line). A soft
cut-off parameter (set $A$) close to the mass of the $f_1$ implies
that all the contributions related to $f_1$ exchange become
negligibly small. On the other hand, for the parameter values
given by set $B$, the $f_1$ exchange contribution is much larger
than that from $\eta$ exchange. Note, that this large uncertainty
in the cut-off presently cannot be controlled by data and we will
discuss the relevance of the $f_1$ exchange contribution for all
reactions separately throughout this study. For set $B$ the total
cross section for the reaction $\pi^-p\rightarrow a_0^-p$ is
about 0.5 mb at 2.4 GeV$/c$ (cf. Fig.~\ref{stot_pip} (upper part))
while the forward differential cross section is about 1
mb/GeV$^2$.

The $\eta$ and $f_1$ exchanges, however, do not contribute to the
amplitude of the charge exchange reaction $\pi^-p\rightarrow a_0^0n$.
In this case we have to consider the contributions of the $s$- and
$u$-channel diagrams (Fig. \ref{Fig1} {\it c} and  \ref{Fig1} {\it d}):
\begin{eqnarray}
&&\hspace*{-4mm}M_N^s(\pi^-p\to a_0^0n) = g_{a_0NN} {f_{\pi NN}\over m_\pi}
\ {1\over s-m_N^2} F_N(s) \times
\nonumber \\
&\times&p_{1\mu}\ \bar u(p_2^\prime)  \left[(p_1+p_2)_\alpha \gamma_\alpha
+m_N\right] \gamma_\mu \ \gamma_5 u(p_2);
\label{eqpip1}
\end{eqnarray}
\begin{eqnarray}
&&\hspace*{-4mm}M_N^u(\pi^-p\to a_0^0n) = g_{a_0NN} {f_{\pi NN}\over m_\pi} \
{1\over u-m_N^2} F_N(u)  \times \nonumber\\
&\times&p_{1\mu} \ \bar u(p_2^\prime) \gamma_\mu \gamma_5
\left[(p_2-p_1^\prime)_\alpha \gamma_\alpha + m_N\right] u(p_2),
\label{eqpip2}
\end{eqnarray}
where $s=(p_1+p_2)^2, \ u=(p_2-p_1^\prime)^2$ and  $m_N$ is the
nucleon mass.

The $\pi NN$ coupling constant is taken as $f_{\pi NN}^2/4\pi
=0.08$~\cite{Holinde} and the form factor for each virtual nucleon
is taken in the so-called monopole form
\begin{eqnarray}
F_N(u) = \frac{\Lambda_N^4}{\Lambda_N^4+(u-m_N^2)^2}.
\label{FN}\end{eqnarray}
Following \cite{Grishina3} we adopt here a cut-off
parameter $\Lambda_N =1.24$~GeV (see also discussion below).

The the rare-dotted and dash-double-dotted lines in the lower part of
Fig.~\ref{dsdt_pip} show the differential cross section for the charge
exchange reaction $\pi^-p\rightarrow a_0^0n$ at 2.4 GeV$/c$
corresponding to $s$- and $u$-channel diagrams, respectively. Due to
isospin constraints  only the $s$ channel contributes to the
$\pi^-p\to a_0^-p$ reaction (rare-dotted line in the upper part of
Fig.~\ref{dsdt_pip}). In these calculations the cut-off parameter
$\Lambda_N$ = 1.24 GeV and $g_{a_0NN}^2/4\pi \simeq 1$ have been
employed in line with the Bonn potential \cite{Holinde}.  The solid
line in the lower part of Fig.~\ref{dsdt_pip} describes the coherent
sum of the $s$- and $u$-channel contributions. Except for the very
forward region the $s$-channel contribution (rare-dotted line) is rather
small compared to the $u$ channel for the charge exchange reaction
$\pi^-p\rightarrow a_0^0n$, which may give a backward differential
cross section of about 1 mb/GeV$^2$ . The corresponding total cross
section can be about 0.3 mb at this energy (cf.  Fig.~\ref{stot_pip},
middle part).

There is a single experimental point for the forward differential cross
section of the reaction $\pi^-p\rightarrow a_0^0n$ at 2.4 GeV$/c$
(\cite{Cheshire}, lower part of Fig. \ref{dsdt_pip}),
$$\left.{d\sigma\over dt}(\pi^-p\rightarrow a_0^0n)\right|_{t\approx 0}
= 0.49 \ \rm{mb/GeV}^2.$$
Since in the forward region ($t \approx$ 0) the $s$-
and $u$-channel diagrams only give a smaller cross section, the charge
exchange reaction $\pi^-p\rightarrow a_0^0n$ is most probably dominated
at small $t$ by the isovector $b_1 (1^{+-})$- and $\rho_2 (2^{--})$-
meson exchanges (see, e.g., \cite{Achasov}). Though the couplings
of these mesons to $\pi a_0$ and $NN$ are not known, we can estimate
$\frac{d\sigma}{dt}(\pi^-p\rightarrow a_0^0n)$ in the forward
region using the Regge-pole model as developed by Achasov and
Shestakov \cite{Achasov2}. Note, that the Regge-pole model is
expected to provide a reasonable estimate for the cross section at
medium energies of about a few GeV and higher (see, e.g.,
\cite{Kaidalov1,Kondrat} and references therein).

\subsection{The Regge-Pole Model}

   The $s$-channel helicity amplitudes for the reaction $\pi^-p
\rightarrow a_0^0n$ can be written as
\begin{eqnarray}
&&\hspace*{-4mm}M_{\lambda_2^\prime\lambda_2}(\pi^-p\rightarrow a_0^0n) =
\bar u_{\lambda_2^\prime}(p_2^\prime)\ \left[-A(s,t) \right. + \nonumber\\
&+&\left.(p_1+p_1^\prime)_\alpha \gamma_\alpha {B(s,t)\over 2}\right]
\gamma_5 u_{\lambda_2}(p_2),
\label{Reg1}\end{eqnarray}
where the invariant amplitudes $A(s,t)$ and $B(s,t)$ do not contain
kinematical singularities and (at fixed $t$ and large $s$) are related
to the helicity amplitudes as
\begin{eqnarray}
M_{++}\approx -sB, \hspace{3mm} M_{+-}\approx \sqrt{t_{\min }-t}\ A. \label{Reg2}\end{eqnarray}
The differential
cross section then can be expressed through the helicity
amplitudes in the standard way as
\begin{eqnarray}
\hspace*{-4mm} {d\sigma\over dt}(\pi^-p\rightarrow a_0^0n)
={1\over 64 \pi s} {1\over (p_1^{\rm{cm}})^2}  (|M_{++}|^2+|M_{+-}|^2).
\label{eq:sigQGSM1}
\end{eqnarray}
Usually it is assumed that the reaction $\pi^-p\rightarrow a_0^0n$
at high energies is dominated by the $b_1$ Regge-pole exchange.
However, as shown by Achasov and Shestakov \cite{Achasov2} this
assumption is not compatible with the angular dependence of
$d\sigma/dt(\pi^-p \rightarrow a_0^0n)$ observed at Serpukhov at
40 GeV$/c$ \cite{Serpukhov,Serpukhov1} and Brookhaven at 18 GeV$/c$
\cite{Brookhaven}. The reason is that the $b_1$ Regge trajectory
contributes only to the amplitude $A(s,t)$ giving a dip in
differential cross section at forward angles, while the data show
a clear forward peak in $d\sigma/dt(\pi^-p\rightarrow a_0^0n)$ at
both energies. To interpret this phenomenon Achasov and Shestakov
introduced a $\rho_2$ Regge-pole exchange conspiring with its
daughter trajectory.  Since the $\rho_2$ Regge trajectory
contributes to both invariant amplitudes, $A(s,t)$ and $B(s,t)$,
its contribution does not vanish at the forward scattering angle
$\Theta =0$ thus giving a
forward peak due to the term $|M_{++}|^2$ in $d\sigma/dt$. At the
same time the contribution of the $\rho _2$ daughter trajectory to
the amplitude $A(s,t)$ is necessary to cancel the kinematical pole
at $t=0$ introduced by the $\rho_2$ main trajectory (conspiracy
effect). In this model the $s$-channel helicity amplitudes can be
expressed through the $b_1$ and the conspiring $\rho_2$ Regge
trajectories exchange as
\begin{eqnarray}
M_{++}\approx M_{++}^{\rho_2}(s,t) = \gamma_{\rho_2}(t)
\exp [-i {\pi\over 2} \alpha_{\rho _2}(t)]
\left( \frac s{s_0}\right)^{\alpha_{\rho_2}(t)}\hspace*{-6mm},
\hspace*{6mm} \label{Reg3}\end{eqnarray}
\begin{eqnarray}
M_{+-}\approx M_{+-}^{b_1}(s,t) &=& \sqrt{(t_{\min }-t)/s_0}\
\gamma_{b_1}(t) \times \nonumber\\
&\times& i\exp [-i {\pi\over 2} \alpha_{b_1}(t)]
\left(\frac s{s_0}\right)^{\alpha_{b_1}(t)},
\label{Reg4}\end{eqnarray}
where $\gamma _{\rho_2}(t) = \gamma_{\rho _2}(0)\ \exp (b_{\rho_2}t)$,
$\gamma_{b_1}(t)\ = \gamma_{b_1}(0)\ \exp (b_{b_1}t)$,\\
$t_{\min}\approx -m_N^2 (m_{a_0}^2-m_\pi ^2)/s^2$,
$s_0\approx 1$ GeV$^2$ while the meson Regge trajectories have the
linear form $\alpha_j(t) = \alpha_j(0)+\alpha_j^\prime(0)t$.

Achasov and Shestakov describe the Brookhaven data on the $t$
distribution at 18~GeV$/c$ for $-t_{\min }\leq -t\leq 0.6$ GeV$^2$
\cite{Brookhaven} by the expression
\begin{eqnarray}
\frac{dN}{dt} = C_1 \left[e^{\Lambda_1t} + (t_{\min }-t)
\frac{C_2}{C_1} e^{\Lambda_2t} \right], \label{Reg5}\end{eqnarray}
where the first and second terms describe the $\rho _2$ and $b_1$
exchanges, respectively. They found two fits: a) $\Lambda_1=4.7$
GeV$^{-2}, C_2/C_1=0,C_1\approx 0$; b) $\Lambda_1=7.6$ GeV$^{-2},
C_2/C_1\approx 2.6$~GeV$^{-2}, \Lambda _2=5.8$~GeV$^{-2}.$ This
implies that at 18 GeV$/c$ the $b_1$ contribution yields only 1/3 of
the integrated cross section. Moreover, using the available data
on the reaction $\pi^-p\rightarrow a_2^0(1320)n$ at 18 GeV$/c$ and
comparing with the data on the $\pi^-p\rightarrow a_0^0n$
reaction they estimated the total and forward differential cross
sections $\sigma (\pi^-p\rightarrow a_0^0n\rightarrow \pi ^0\eta
n)\approx 200$ nb and $[d\sigma /dt(\pi^-p\rightarrow
a_0^0n\rightarrow \pi^0\eta n)]_{t=0}\approx 940$ nb/GeV$^2.$
Taking ${\rm Br}(a_0^0\rightarrow \pi^0\eta)\approx 0.8$ we find $\sigma
(\pi^-p\rightarrow a_0^0n)\approx 0.25$ $\mu$b and $[d\sigma
/dt(\pi^-p\rightarrow a_0^0n)]_{t=0}\approx 1.2$ $\mu$b/GeV$^2$.

In this way all the parameters of the Regge model can be fixed and
we will employ it for the energy dependence of the
$\pi^-p\rightarrow a_0^0n$ cross section to obtain an estimate at
lower energies, too.

The mass of the $\rho_2(2^{--})$ is expected to be about 1.7 GeV
(see \cite{Kokoski} and references therein) and the slope of the
meson Regge trajectory in the case of light ($u, d$) quarks is 0.9
GeV$^{-2}$  \cite{Kaidalov2}. Therefore, the intercept of the
$\rho_2$ Regge trajectory is $\alpha_{\rho
_2}(0)=2-0.9m_{\rho_2}^2\approx -0.6$. Similarly -- in the case of
the $b_1$ trajectory -- we have $\alpha_{b_1}(0)\approx -0.37$. At
forward angles we can neglect the contribution of the $b_1$
exchange (see discussion above) and write the energy dependence of
the differential cross section in the form
\begin{eqnarray}
\left.{d\sigma_{\rm Regge}\over dt}(\pi^{-}p\rightarrow a_0^0n)\right|_{t=0}
&\approx&\left.\frac{d\sigma_{\rho_2}}{dt}\right|_{t=0} \sim\nonumber\\
&\sim&{1\over (p_1^{\rm c.m.})^2}\left(\frac s{s_0}\right)^{-2.2}\hspace*{-5mm}.
\hspace*{5mm} \label{eq:sigQGSM2}
\end{eqnarray}
This provides the following estimate for the forward differential
cross section at 2.4 GeV$/c$,
\begin{eqnarray}
\left.{d\sigma_{\rm Regge}\over dt} (\pi^-p\rightarrow
a_0^0n)\right|_{t=0} \approx 0.6 \ \rm{ mb/GeV}^2,
\label{eq:sigQGSM3}
\end{eqnarray}
which is in agreement with the experimental data point
\cite{Cheshire} (lower part of Fig. \ref{dsdt_pip}). Since the $b_1$ and
$\rho_2$ Regge trajectories have isospin 1, their contribution to
the cross section for the reaction $\pi^-p\rightarrow a_0^-p$ is
twice smaller,
\begin{eqnarray}
{d\sigma_{\rm Regge}\over dt} (\pi^-p\rightarrow a_0^-p)=
\frac 12\ {d\sigma_{\rm Regge}\over dt} (\pi^-p\rightarrow a_0^0n).
\label{eq:sigQGSM4}
\end{eqnarray}
In Fig. \ref{dsdt_pip} the dotted lines  show the resulting
differential cross sections for $d\sigma_{\rm Regge}(\pi^-p\rightarrow
a_0^-p)/dt$ (upper part) and $d\sigma_{\rm Regge}(\pi^-p\rightarrow
a_0^0n)/dt$ (lower part) at 2.4 GeV$/c$ corresponding to $\rho_2$ Regge
exchange, whereas the dash-dotted lines indicate the contribution for
$\rho_2$ and $b_1$ Regge trajectories.  For $t \to 0$ both Regge
parametrizations agree, however, at large $|t|$ the solution including
the $b_1$ exchange gives a smaller cross section. The cross section
$d\sigma_{\rm Regge}(\pi^-p\rightarrow a_0^-p)/dt$ in the forward region
exceeds the contributions of $\eta$, $f_1$ (set $A$) and $s$-channel
exchanges, however, is a few times smaller than the $f_1$-exchange
contribution for set $B$. On the other hand, the cross section
$d\sigma_{\rm Regge} (\pi^-p\rightarrow a_0^0n)/dt$ is much larger than the
$s$- and $u$-channel contributions in the forward region, but much
smaller than the $u$-channel contribution in the backward region.

\begin{figure}[h]
\centerline{\psfig{figure=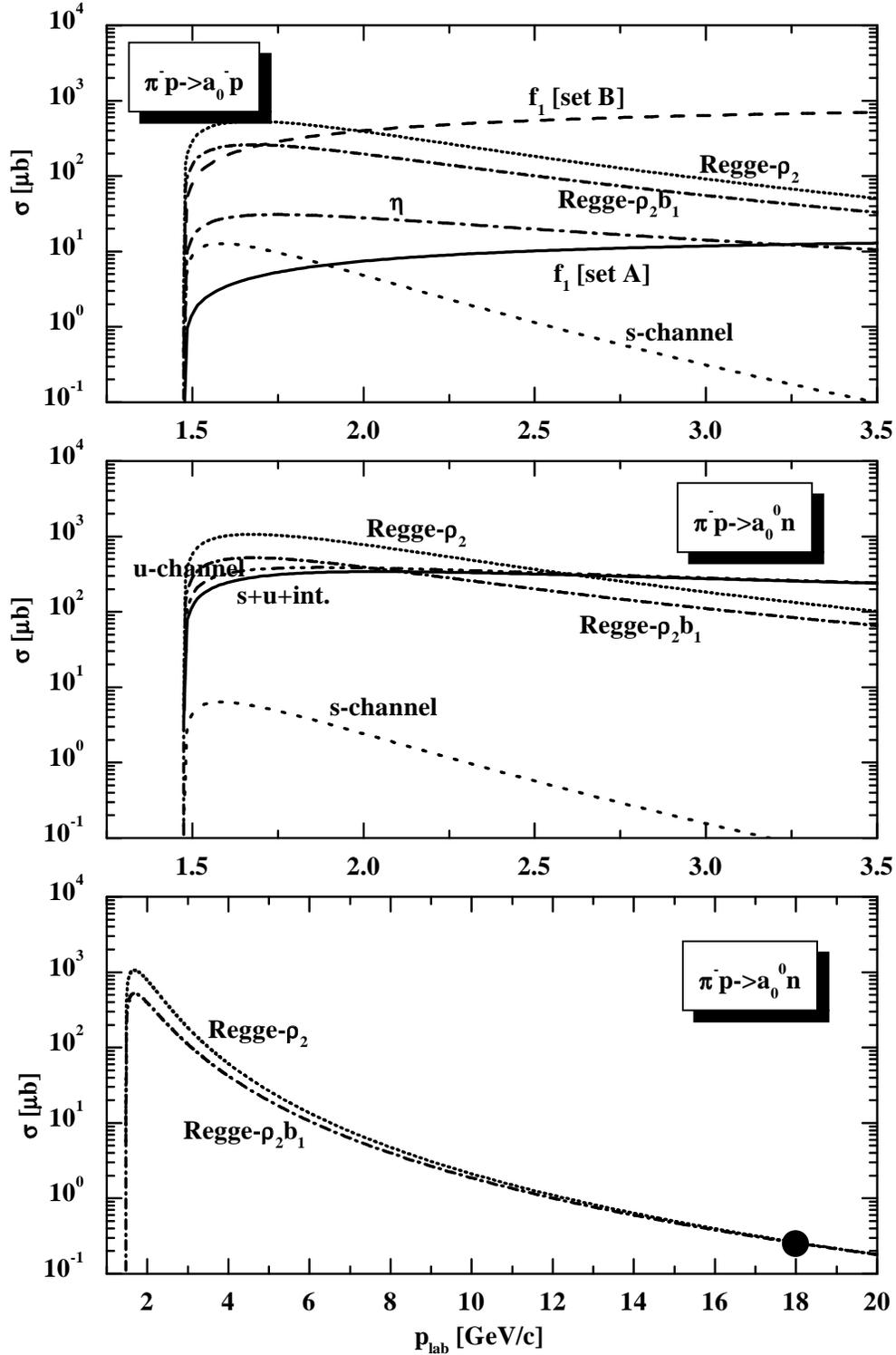,width=13cm}}
\caption{
The total cross sections for the reactions $\pi^-p\rightarrow
a_0^-p$ (upper part) and $\pi^-p\rightarrow a_0^0n$ (middle and lower
part) as a function of the incident momentum. The assignment of the
lines is the same as in Fig. 2.  The experimental data point at 18 GeV$/c$
(lower part) is taken from \protect\cite{Brookhaven}.
}\label{stot_pip}
\end{figure}

The integrated cross sections for $\pi^- p \rightarrow a_0^- p$ (upper
part) and $\pi^- p \rightarrow a_0^0 n$ (middle and lower part) for the
Regge model are shown in Fig. \ref{stot_pip} as a function of the pion
lab. momentum by dotted lines for $\rho_2$ exchange and by
dash-dotted lines for $\rho_2, b_1$ trajectories. In the few GeV region
the cross sections are comparable with the $u$-channel contribution. At
higher energies the Regge cross section decreases as $s^{-3.2}$ in
contrast to the non-Reggeized  $f_1$-exchange contribution which
increases with energy and seems to be too large at 2.5 GeV$/c$ for
parameters from the set $B$. We thus expect parameter set $B$ to be
unrealistic.

The main conclusions of this Subsection are as follows. In the region of a
few GeV the dominant mechanisms of  $a_0$ production in the reaction
$\pi N \rightarrow a_0 N$ is the  $u$-channel nucleon exchange (cf.
middle part of Fig.~\ref{stot_pip}).  Similar cross section ($\simeq$
0.4--1 mb) is predicted by the Regge model with conspiring $\rho_2$
(or $\rho_2$ and $b_1$) exchanges, normalized to the Brookhaven data at
18 GeV$/c$ (lower part of Fig.~\ref{stot_pip}). The contributions of
$s$-channel nucleon and $t$-channel $\eta$-meson exchanges are small
(cf. upper and middle parts of Fig.~\ref{stot_pip}).

\subsection{Possible Signals of $a_0$ Production in the Reaction\\
$\pi N \to K \bar K N$}

\begin{figure}[h]
\centerline{\psfig{figure=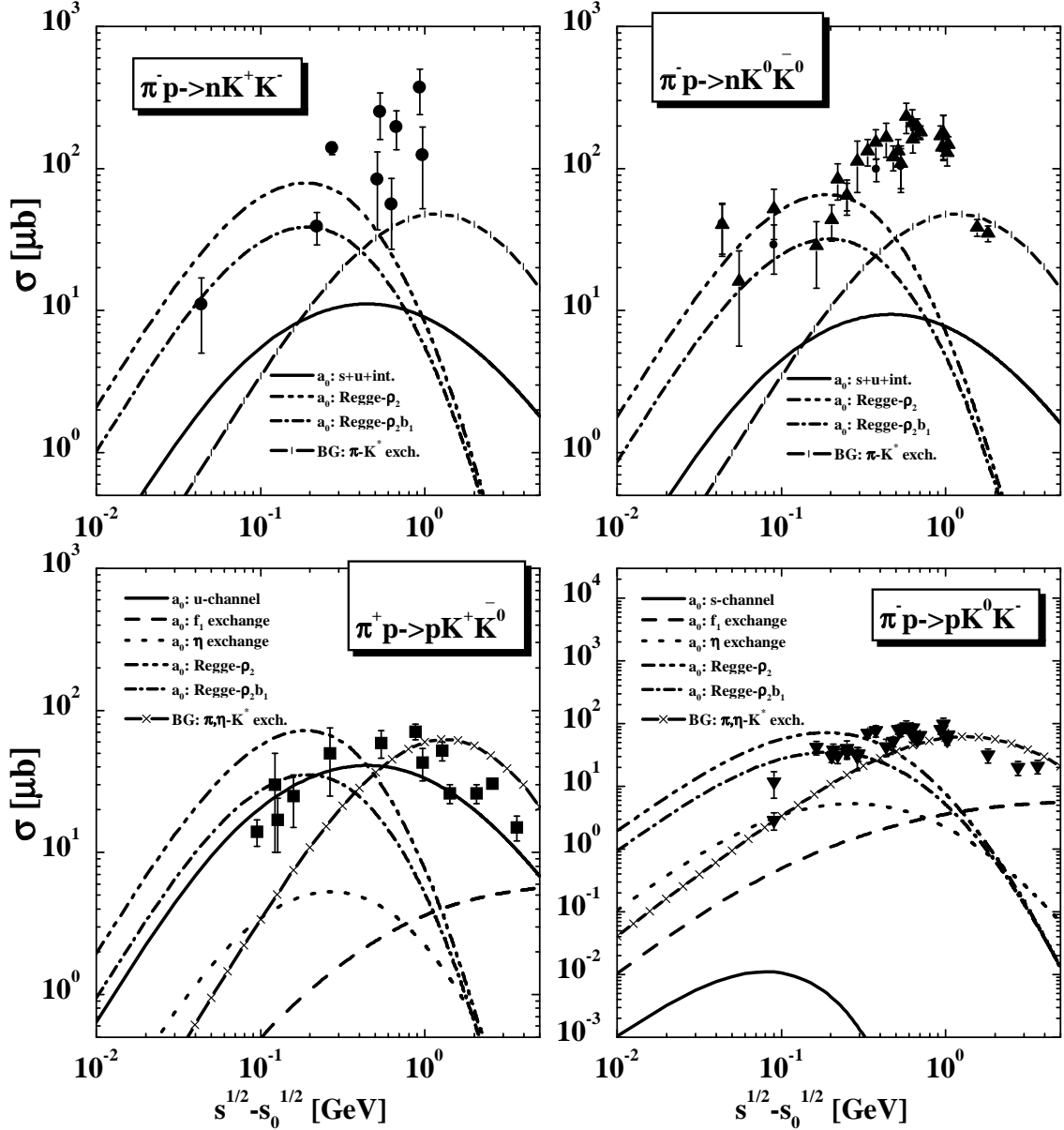,width=15cm}}
\vspace*{1cm}\caption{
The total cross sections for the reactions $\pi^- p \to n K^+
K^-$ (upper left), $\pi^- p \to nK^0  \bar {K^0}$ (upper right), $\pi^+
p \to p K^+ \bar {K^0} $ (lower left) and $\pi^- p \to pK^0 K^- $
(lower right). Experimental data are taken from \cite{Landolt}.
The solid curves describe $s$- and $u$-channel contributions, calculated
with the dipole nucleon form factor $(F^2_N(u)$ with $\Lambda_N =
1.35$ GeV.  The short-dashed and long-dashed curves describe $\eta$ and
$f_1$ $t$-channel exchanges, respectively. Two different choices of
the Regge-pole model are shown by the dash-dotted curves which describe
$\rho_2$-exchange (upper) and conspiring $\rho_2 b_1$ -exchange
(lower). The crossed solid lines show the background contribution from
diagram e) in Fig.~1.
}\label{pinkk_bg}
\end{figure}

   In Fig. \ref{pinkk_bg} we show the existing experimental data on the
reactions $\pi^- p \to n K^+ K^-$ (upper left), $\pi^- p \to nK^0  \bar
{K^0}$ (upper right), $\pi^+ p \to p K^+ \bar {K^0} $ (lower left), and
$\pi^- p \to pK^0 K^- $ (lower right) taken from \cite{Landolt}.
The solid curves describe s- and u-channel contributions, calculated
using the dipole nucleon form factor $(F^2_N(u))$ with $\Lambda_N =
1.35$ GeV.  The short-dashed and long-dashed curves describe $\eta$ and
$f_1$ $t$-channel exchanges, respectively. Two different choices of
the Regge-pole model are shown by the dash-dotted curves which describe
$\rho_2$ exchange (upper) and $\rho_2 b_1$ exchange (lower). The
crossed solid lines display the background contribution (see diagram e) in
Fig. \ref{Fig1}) which was calculated using parameters of the $K^*$
exchange from the J\"ulich model \cite{Jan}. It is important that for
the reactions $\pi^+ p \to p K^+ \bar {K^0} $ and $\pi^- p \to pK^0 K^-
$, where the $K\bar K$ pair has isospin 1, the main contributions
come from $P$-wave $K\bar K$ pair production from the $\pi \pi$ state and
from $S$-wave $K \bar K$ pair production from the $\eta \pi$ state.
These selection rules follow from $G$-parity conservation (note that
the $G$ parity of the $K \bar K$ system with orbital momentum $L$ and
isospin $I$ is given by $(-1)^{L+I}$).  At the same time for the
reactions $\pi^- p \to n K^+ K^-$ and $\pi^- p \to nK^0  \bar {K^0}$
the essential contribution to the background stems from $S$-wave
$K\bar K$ pair production from the isoscalar $\pi \pi$ state.  Let us
note that the parametrization of the total cross sections for the reactions
$\pi N \to K \bar K N$ has been discussed previously in
\cite{Sibirtsev1}. Here we analyze also contributions from
different channels to the total cross sections.

The most important point is that for all the reactions the
background  is essentially below the data at the c.m. energy release
$Q \leq 300$ MeV. In
case of the reactions $\pi^+ p \to p K^+ \bar {K^0} $ and $\pi^- p
\to pK^0 K^- $ this, to our opinion, can only be due to a
contribution of the $a_0$. Of course, in the reactions $\pi^- p \to n K^+
K^-$ and $\pi^- p \to nK^0  \bar {K^0}$ both scalar mesons, $f_0$ and
$a_0$, can contribute. In a series of bubble chamber experiments,
performed in 60$-$70-ties, a structure was reported in the mass
distribution of the $K_s^0 K_s^0$ system produced in the reaction
$\pi^- p \to nK^0_s {K^0_s}$ (see, e.g., \cite{Dahl} and references
therein).  Usually this structure was attributed to the $f_0(980)$. In
our previous work we used the data on $\pi^- p \to n f_0 \to nK^0_s
{K^0_s}$ to find a restriction on the branching ${\rm Br}(f_0 \to K \bar K$)
\cite{Brat99}. We see here from Fig. \ref{pinkk_bg} (upper right) that
an important contribution to the cross section of the reaction
$\pi^- p \to nK^0 \bar {K^0}$ at $Q \leq 300$ MeV comes also from the
$a_0$.  We cannot exclude that there can also be some contribution from
$a_0(980)$ at $Q \geq 300$ MeV. If this is really the case, our
restriction on ${\rm Br}(f_0 \to K \bar K)$ \cite{Brat99} has to be
corrected. This problem, however, requires further analysis.

Let us note that the amplitude corresponding to the Feynman diagram {\it e})
in Fig. \ref{Fig1}  would predict a sharply rising cross section for $Q
\geq 400$ MeV. To suppress this unrealistic behavior we used a
Reggeized $K^*$- propagator multiplying the Feynman propagator of the
vector meson in all the amplitudes by the Regge power
$(s/s_0)^{(\alpha_{K^*}(0)-1)}$ with $\alpha_{K^*}(0) \simeq 0.25$,
$\sqrt{s_0}= 2m_K +m_N$.  The background curves are in reasonable
agreement with the data on the reactions $\pi^+ p \to p K^+ \bar {K^0}
$ and $\pi^- p \to pK^0 K^- $ at $Q \geq 400$ MeV (see the crossed
solid lines in two lower parts of Fig. \ref{pinkk_bg}).

The Regge-pole model for $a_0$ production, especially the set with
$b_1\rho_2$ exchange, is in a good agreement with the data for all the
reactions at $Q \leq$300 MeV giving a cross section of the reaction
$\pi N \to a_0 N \to K \bar K N$ of about 20$-$30 $\mu$b at $Q \simeq
100$--$300$ MeV. At larger $Q$ it drops very fast. The $u$-channel
contribution is also in a good agreement with the data on the reaction
$\pi^+ p \to p K^+ \bar {K^0}$, but the coherent sum of the $u$-
and $s$-channel contributions is below the data for the
reactions $\pi^- p \to n K^+ K^-$ and $\pi^- p \to nK^0  \bar {K^0}$.
The $t$-channel $\eta$ and $f_1$ exchange contributions
are small and can be neglected.

Note that both invariant mass distributions of the $K^- \bar{K^0}$ and
$K^0_s {K^0_s}$ systems presented in \cite{Dahl} show a resonance-like
structure near the $K \bar K$ threshold at $Q \leq 300$ MeV.  However,
because of a comparatively small number of events for each fixed
initial momentum those distributions are averaged over a large interval
of about 1 GeV$/c$ in $p_{\rm{lab}}$. Unfortunately, those
distributions cannot be directly compared with theoretical ones at any
fixed $Q$ especially in the near-threshold region.  In order to give
another strong argument, that the $a_0$ contribution is really
necessary to explain the existing experimental data, let us consider
the energy dependence of the total cross section of the reaction
$\pi^- p \to p K^- \bar {K^0} $.  Averaging the existing data from
\cite{Landolt} versus $p_{{\rm lab}}$ over the intervals $2.0 \pm 0.15$
and $3.0 \pm 0.15$ GeV$/c$ we find $\sigma_{\rm av}= 34.9 \pm 3.3$ and $
73.8\pm 7.6 ~\mu$b, respectively.  The ratio of those cross sections
is equal to $R_{21} \simeq 2.1 \pm 0.05$. The energy behaviour of the
background contribution in our model is $\sigma_{{\rm bg}} \sim
Q^{2.3}$. If we assume that in the interval of $Q =250 - 630$ MeV (which corresponds to the interval of $p_{\rm lab}=2$--3 GeV/$c$) the
background contribution is present only, we get $R^{{\rm bg}}_{21}
\simeq 5.5$. This means that at 3 GeV$/c$ we should expect cross
section $\simeq 200$~$\mu$b instead of  $\sim 70\ \mu$b. Evidently, experimental data are inconsistent with this assumption.

Let us formulate the main conclusions of this Subsection. The existing
data on the reactions $\pi^+ p \to p K^+ \bar {K^0} $ and $\pi^- p \to
pK^0 K^- $ give a rather strong evidence that at low energy above
threshold ($Q \leq 300$ MeV) they are dominated by $a_0$
production. The same is true also for the reactions $\pi^- p \to n K^+ K^-$
and $\pi^- p \to nK^0  \bar {K^0}$, where some smaller contribution of
$f_0$ may also be present.  The value of the $a_0$ production cross
section is reasonably described by the Regge-pole model with
($\rho_2,b_1$) exchange as proposed by Achasov and Shestakov
\cite{Achasov2}. The $u$-channel exchange mechanism also gives
a reasonable value of the cross section.

\section{The Reaction $N N\to N N a_0$}

\subsection{An Effective Lagrangian Approach with One-Pion Exchange}

\begin{figure}[h]
\centerline{\psfig{figure=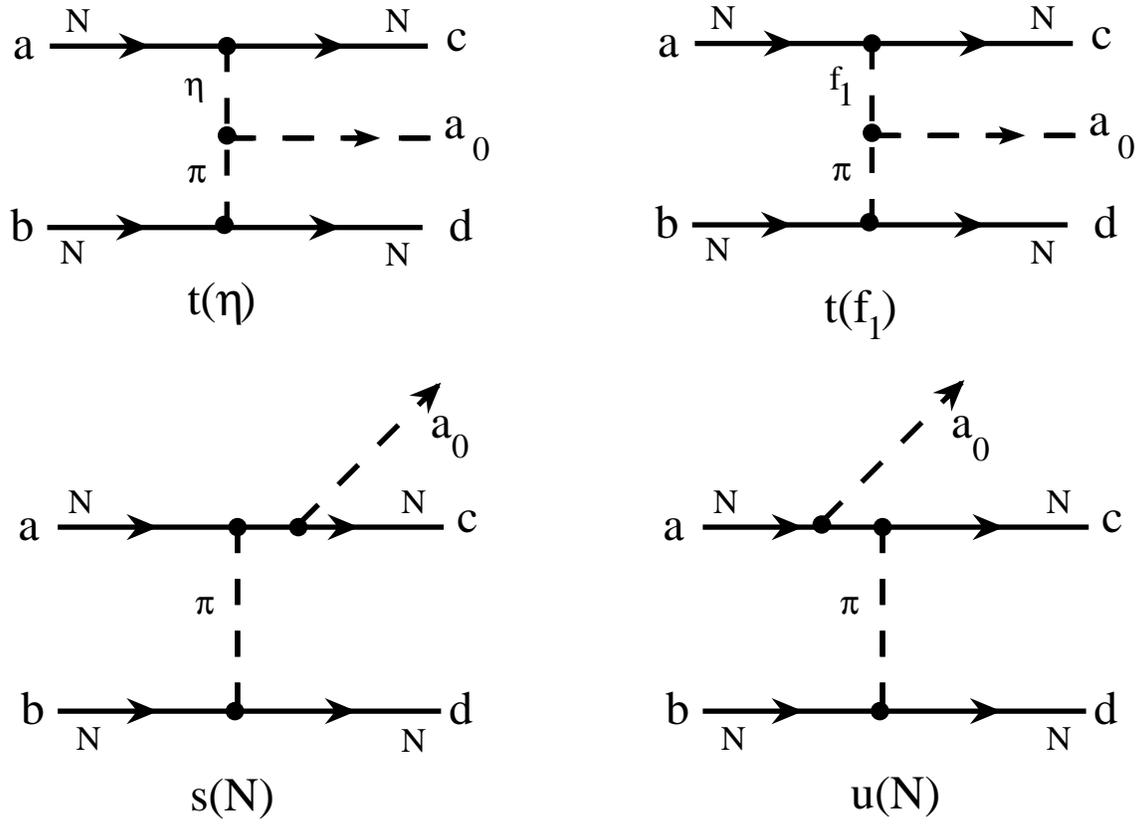,width=15cm}}
\vspace*{1cm}\caption{
Diagrams for $a_0$ production in the reaction $N N\rightarrow
a_0 N N$.
}\label{diagr_a0}
\end{figure}

   We consider $a_0^0$, $a_0^+$, $a_0^-$ production in the reactions
$j=pp\to pp a_0^0$, $pp\to pn a_0^+$, $pn\to pp a_0^-$, and $pn\to pn
a_0^0$ using the effective Lagrangian approach with one-pion exchange
(OPE). For the elementary $\pi N\to N a_0$ transition amplitude we take
into account different mechanisms $\alpha$ corresponding to $t$-channel
diagrams with $\eta(550)$- and $f_1(1285)$-meson exchanges
($\alpha=t(\eta)$, $t(f_1)$) as well as $s$- and $u$-channel graphs
with an intermediate nucleon ($\alpha=s(N)$, $u(N)$) (cf.
\cite{Grishina3}).  The corresponding diagrams are shown in
Fig.~\ref{diagr_a0}.  The invariant amplitude of the $NN\to NN a_0$
reaction then is the sum of the four basic terms (diagrams in Fig.
\ref{diagr_a0}) with permutations of nucleons in the initial and
final states
\begin{eqnarray}
\hspace*{-5mm} {\mathcal{M}}_{j(\alpha)}^{\pi}[ab;cd]
&&=\xi^{\pi}_{j(\alpha)}[ab;cd] \ {\mathcal{M}}_{\alpha}^{\pi}[ab;cd] +
\xi^{\pi}_{j(\alpha)}[ab;dc] \ {\mathcal{M}}_{\alpha}^{\pi}[ab;dc]+
\label{NNa0sum} \\
&&+\xi^{\pi}_{j(\alpha)}[ba;dc] \ {\mathcal{M}}_{\alpha}^{\pi}[ba;dc] +
\xi^{\pi}_{j(\alpha)}[ba;cd] \ {\mathcal{M}}_{\alpha}^{\pi}[ba;cd],
\nonumber
\end{eqnarray}
where the coefficients $\xi^{\pi}_{j(\alpha)}$ are given in
Table.  The amplitudes for the $t$-channel exchange with
$\eta(550)$- and  $f_1(1285)$-mesons are given by
\begin{eqnarray}
{\mathcal{M}}_{t(\eta)}^{\pi}[ab;cd] &=&
g_{a_0\eta\pi} F_{a_0\eta\pi}\left((p_a-p_c)^2,(p_d-p_b)^2\right)
\  g_{\eta NN}
F_{\eta }\left((p_a-p_c)^2\right)
 \times \nonumber \\
&\times&  {1\over (p_a-p_c)^2-m_\eta ^2} \
\bar u(p_c) \gamma_5 u(p_a)\times {\mathrm{\Pi}}(p_b;p_d),
\label{NN-eta}
\end{eqnarray}
\begin{eqnarray}
{\mathcal{M}}_{t(f_1)}^{\pi}[ab;cd] &=&
-g_{a_0 f_1\pi} F_{a_0 f_1\pi} \left((p_a-p_c)^2,(p_d-p_b)^2\right)
g_{f_1 NN} F_{f_1}\left((p_a-p_c)^2\right) \times \nonumber\\
&\times& {1\over (p_a-p_c)^2-m_{f_1}^2}
 \ (p_a-p_c+2 \ (p_b-p_d))_\mu   \times \nonumber\\
&\times& \left(g_{\mu\nu}-{(p_a-p_c)_\mu (p_a-p_c)_\nu\over
        m_{f_1}^2}\right) \times \nonumber\\
&\times& \bar u(p_c) \gamma_5\gamma_{\nu} u(p_a)\times
 {\mathrm{\Pi}}(p_b;p_d),
\label{NN-f1}
\end{eqnarray}
with
\begin{eqnarray}
{\mathrm{\Pi}}(p_b;p_d) &=&
\frac{f_{\pi NN}}{m_{\pi}}\ F_{\pi }\left((p_b-p_d)^2\right)
(p_b-p_d)_{\beta} \
\bar u(p_d) \gamma_5 \gamma_{\beta} u(p_b) \times \nonumber\\
&\times&  \frac{1}{(p_b-p_d)^2-m_{\pi}^2}.
\label{piNN}
\end{eqnarray}
The amplitudes for the $s$ and $u$ channels (lower part of
Fig.~\ref{diagr_a0}) are given as
\begin{eqnarray}
{\mathcal{M}}_{s(N)}^{\pi}[ab;cd] &=&
{\mathrm{\Pi}}(p_b;p_d)\
\frac{f_{\pi NN}}{m_{\pi}} F_{\pi }\left((p_d-p_b)^2\right)
\ g_{a_0 NN} \times \label{NN-s}\\
&\times&{ F_{N} \left((p_a+p_b-p_d)^2\right)\over
        (p_a+p_b-p_d)^2-m_N^2} \times \nonumber\\
&\times& (p_d-p_b)_{\mu}\
 \bar u(p_c)[(p_a+p_b-p_d)_{\delta}\gamma_{\delta}+m_N]
\gamma_5 \gamma_{\mu} u(p_a), \nonumber
\end{eqnarray}
\begin{eqnarray}
{\mathcal{M}}_{u(N)}^{\pi}[ab;cd] &=&
{\mathrm{\Pi}}(p_b;p_d)\
\frac{f_{\pi NN}}{m_{\pi}} F_{\pi }\left((p_d-p_b)^2\right)
\ g_{a_0 NN} \times \label{NN-u}\\
&\times& {F_{N}\left((p_c+p_d-p_b)^2\right)
\over (p_c+p_d-p_b)^2-m_N^2} \times \nonumber\\
&\times& (p_d-p_b)_{\mu}
\ \bar u(p_c)\gamma_5 \gamma_{\mu} [(p_c+p_d-p_b)_{\delta}
\gamma_{\delta}+m_N] u(p_a).
\nonumber
\end{eqnarray}
Here $p_a, p_b$ and $p_c, p_d$ are the four momenta of the initial and
final nucleons, respectively.
As in the previous Section we mostly employ coupling constants and form
factors from the Bonn$ -$J\"ulich potentials (see, e.g.,
\cite{Holinde,Bonnf1,Haidenbauer}).

For the form factors at the $a_0 f_1 \pi$ (as well as $a_0 \eta\pi$) vertex
factorized forms are applied following the assumption from
\cite{Chung,Nakayama},
\begin{eqnarray}
F_{a_0 f_1 \pi}(t_1,t_2)=F_{f_1 NN}(t_1) \ F_{\pi NN}(t_2),
\label{ff_f1pia0}\end{eqnarray}
where $F_{f_1 NN}(t), F_{\pi NN}(t)$ are taken in the monopole form
(see previous Section). Usually the cut-off parameter $\Lambda_{\pi NN}$
is taken in the interval 1$-$1.3 GeV. Here we take
$\Lambda_{\pi NN}=1.05$ GeV (see also the discussion in \cite{Brat01}).

As shown in the analysis of \cite{Grishina3} the contribution of
the $\eta$ exchange to the amplitude $\pi N \to a_0 N$ is small (cf. also
Section 3).  Note that in \cite{Baru2} only this mechanism
was taken into account for the reaction $pn \to pp a_0^-$.  Here we
also include the $\eta$ exchange because it might be noticeable in
those isospin channels where a strong destructive interference of $u$-
and $s$-channel terms can occur (see below).

Since we have two nucleons in the final state it is necessary to take
into account their final state interaction (FSI), which has some
influence on meson production near threshold. For this purpose we adopt
the FSI model from \cite{BaruFSI} based on the (realistic) Paris
potential. We use, however, the enhancement factor $F_{NN}(q_{NN})$ --
as given by this model -- only in the region of small relative momenta
of the final nucleons $q_{NN} \leq q_0$, where it is larger than 1.
Having in mind that this factor is rather uncertain at larger $q_{NN}$,
where for example contributions of nonnucleon intermediate states to
the loop integral might be important, we assume that $F_{NN}(q_{NN})
=1$ for $q_{NN} \geq q_0$.

\begin{figure}[h]
\centerline{\psfig{figure=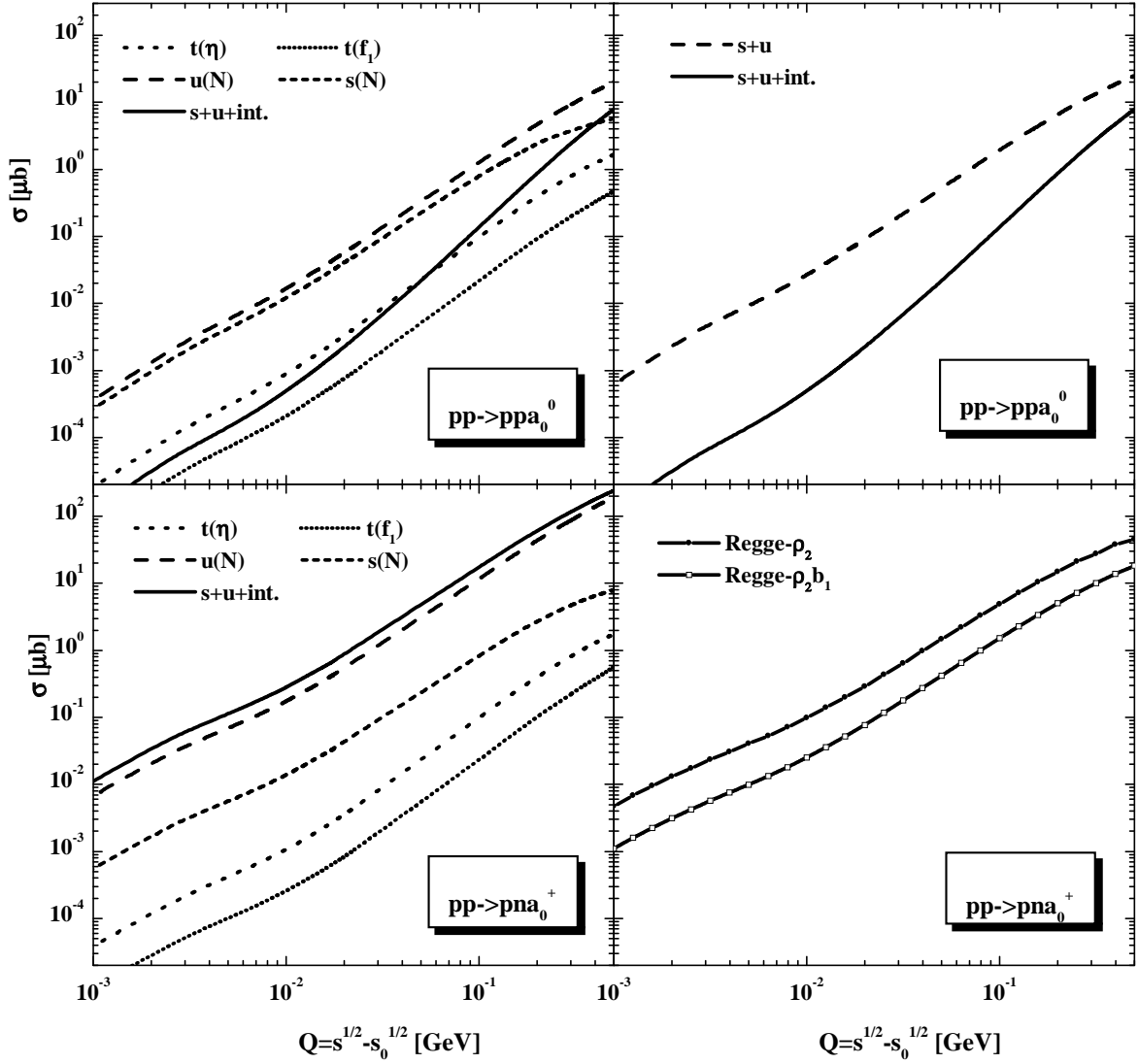,width=15.5cm}}
\vspace*{1cm}\caption{
The total cross sections for the reactions $pp\to pp a_0^0$
(upper part) and $pp\to pn a_0^+$ (lower part) as a function of the
excess energy $Q=\sqrt{s}-\sqrt{s_0}$ calculated with FSI.  The short
dotted lines (l.h.s.) corresponds to the $t(f_1)$ channel, the dotted
lines to the $t(\eta)$ channel, the dashed lines to the $u(N)$ channel,
the short dashed lines to the $s(N)$ channel.  The dashed line (upper
part, r.h.s.) is the incoherent sum of the contributions from $s(N)$
and $u(N)$ channels ($s+u$).  The solid lines indicate the coherent sum
of $s(N)$ and $u(N)$ channels with interference ($s+u+int.$).  The
solid lines with full dots and with open squares (lower part, r.h.s.)
present the results within the $\rho_2$ and $(\rho_2,b_1)$ Regge
exchange model.
}\label{pp_q}
\end{figure}

In Fig.~\ref{pp_q} we show the total cross section as a function of the
energy excess $Q=\sqrt{s}-\sqrt{s_0}$ for the reactions -- $pp\to
pp a_0^0$ (upper part) and $pp\to pn a_0^+$ (lower part).  The solid
lines with full dots and with open squares (r.h.s.) represent the
results within the $\rho_2$ and $(\rho_2,b_1)$ Regge exchange model.
The dotted lines (l.h.s.) correspond to the $t(f_1)$ channel,
the rare-dotted lines to the $t(\eta)$ channel, the dashed lines to the
$u(N)$ channel, the short dashed lines to the $s(N)$ channel.  The
dashed line in the right upper part of Fig.~\ref{pp_q} is the
incoherent sum of the contributions from $s(N)$ and $u(N)$ channels
($s+u$).

As seen from Fig. \ref{pp_q}, the $u$ and $s$ channels give the
dominant contribution; the $t(f_1)$ channel is small for both isospin
reactions.  For the reaction $pp\to pn a_0^+$, the Regge exchange
contribution (extended to low energies) becomes important.  For the $pp
\to pp a_{0}^0$ channel the Regge model predicts no contribution from
$\rho_2$ and ($\rho_2,b_1$) exchanges due to isospin arguments (i.e.,
the vertex with a coupling of three neutral components of isovectors
vanishes); thus only $s$, $u$, $t(\eta)$, and $t(f_1)$ channels are
plotted in the upper part of Fig.~\ref{pp_q}.

Here we have to point out the influence of the interference between the
$s$ and $u$ channels. According to the isospin coefficients from the
OPE model presented in Table, the phase (of interference
$\alpha$) between the $s$ and $u$ channels
${\mathcal{M}}_{s(N)}^{\pi}+\exp(-i\alpha){\mathcal{M}}_{u(N)}^{\pi}$
is equal to zero, i.e., the sign between ${\mathcal{M}}_{s(N)}^{\pi}$
and ${\mathcal{M}}_{u(N)}^{\pi}$ is ``plus''.  The solid lines in
Fig.~\ref{pp_q} indicate the coherent sum of $s(N)$ and
$u(N)$ channels including the interference of the amplitudes
($s+u+$int.). One can see that for the $pp\to pn a_0^+$
reaction the interference is positive and
increases the cross section, whereas for the $pp\to pp a_0^0$ channel
the interference is strongly destructive since we have identical
particles in the initial and final states and the contributions of $s$
and $u$ channels are very similar.

Here we would like to comment  about an extension of the OPE (one-pion
exchange) model to an OBE (one-boson exchange) approximation, i.e.,
accounting for the exchange of $\sigma, \rho, \omega, ...$ mesons as
well as for multi-meson exchanges.  Generally speaking, the total cross
section of $a_0$ production should contain the sum of all the
contributions:
$$\sigma(NN\to NNa_0) = \Sigma_j \sigma_j,$$
where $j=\pi,\sigma,\rho,\omega...$.  Depending on their cut-off
parameters the heavier meson exchanges might give a comparable
contribution to the total cross section for $a_0$ production. An
important point, however, is that near threshold (e.g. $Q \leq 0.3-0.6$
GeV) the energy behavior of all those contributions is the same, i.e.,
it is proportional to the three-body phase space $ \sigma_j \sim Q^2$
(when the FSI is switched off and the narrow resonance width limit is
taken). In this respect we can consider the one-pion exchange as an
effective one and normalize it to the experimental cross section by
choosing an appropriate value of $\Lambda_{\pi}$.  The most appropriate
choice for $\Lambda_{\pi}$ is about 1 -- 1.3 GeV.  Another question
is related to the isospin of the effective exchange.  As it is known
from a serious of papers on the reactions $NN\to NNX,
X=\eta,\eta^{\prime},\omega,\phi$ the  most important
contributions to the corresponding cross sections near threshold come
from $\pi$ and $\rho$ exchanges (see, e.g., the review
\cite{NakayamaReview} and references therein). In line with those
results we assume here that the dominant contribution to the cross
section of the reaction $NN \to NNa_0$ comes also from  the isovector
exchanges (like $\pi$ and $\rho$).  In principle, it is also possible
that some baryon resonances may contribute. However, there is no
information about resonances which couple to the $a_0N$ system.  Our
assumptions thus enable us to make exploratory estimates of the $a_0$
production cross section without introducing free parameters that would
be out of control by existing data. The model can be extended
accordingly when new data on the $a_0$ production will be available.

Another important question is related to the choice of the form factor
for a virtual nucleon, that -- in line with the Bonn$-$J\"ulich
potentials -- we choose as given by (\ref{FN}), which corresponds to
monopole form factors at the vertices.  In the literature, furthermore,
dipole-like form factors (at the vertices) are also often used (cf.
\cite{Nakayama,NakayamaReview,Feuster}).  However,
there are no strict rules for the ``correct'' power of the nucleon form
factor. In physics terms, the actual choice of the power should  not be
relevant; we may have the same predictions for any reasonable choice of
the power if the cut-off parameter $\Lambda_N$ is fixed accordingly.
Note, that $\Lambda_N$ may also depend on the type of mesons involved
at the vertices.
In our previous work \cite{Grishina3} we have fixed $\Lambda_N$ for the
monopole related form factor (\ref{FN})  in the interval 1.2--1.3 GeV
fitting the forward differential cross section of the reaction $pp \to
da_0^+$ from \cite{BNL73}. On the other hand, the same data can be
described rather well using a dipole form factor (at the vertices) with
$\Lambda_N=$1.55$-$1.6 GeV. If we employ this
dipole form factor with $\Lambda_N=$1.55--1.6 GeV in the present case we
obtain practically identical predictions for the cross sections of the
channels $pp \to pn a_0^+$,  $pn \to pn a_0^0$, $pn \to pp a_0^-$,
where the $u$-channel mechanism is dominant and $u-s$ interference is
not too important. In the case of the channel $pp \to pp a_0^0$ we
obtain cross sections by up to a factor of 2 larger for the dipole-like
form factor in comparison to the monopole one.  This is related to the
strong destructive interference of the $s$ and $u$ exchange mechanisms,
which slightly depends on the type of form factor used.  However, our
central result, that the cross section for the $pn a_0^+$ final channel
is about an order of magnitude higher than the $ppa_0^0$ channel in
$pp$ collisions, is robust (within less than a factor of 2) with
respect to different choices of the form factor.

As seen from Fig.~\ref{pp_q}, we get the largest cross
section for the $pp\to pn a_0^+$ isospin channel. For this reaction the
$u$ channel gives the dominant contribution, the $s$-channel cross
section is small such that the interference is not so essential as for
the $pp\to pp a_0^0$ reaction.

As it was already discussed in our previous study \cite{Grishina3} an
effective Lagrangian model cannot be extrapolated to high energies
because it predicts the elementary amplitude $\pi N \to a_0N$ to rise
fast. Therefore, such model can only be employed not far from the
threshold.  On the other hand, the Regge model is valid at large
energies and we have to worry, how close to the threshold we can
extrapolate corresponding amplitudes.  According to duality arguments
one can expect that the Regge amplitude can be applied at low energy,
too, if the reaction $\pi N \to a_0N$ does not contain essential
$s$-channel resonance contributions. In this case the Regge model might
give a realistic estimate of the $\pi N \to a_0N$ and $NN \to NNa_0$
amplitudes even near threshold.

Anyway, as we have shown in \cite{Grishina3} (see
also Section 3) the Regge and $u$-channel model give quite
similar results for the $\pi^- p \to a_0^0 n$ cross section in the near
threshold region; some differences in the cross sections of the
reactions $NN \to NNa_0$ -- as predicted by those two models -- can be
attributed to differences in the isospin factors and effects of $NN$
antisymmetrization which is important near threshold (the latter was
ignored in the Regge model formulated for larger energies).

\subsection{The Reaction $N N\to NN a_0 \to N N K \bar K$}

\subsubsection{Numerical Results for the Total Cross Section}

\begin{figure}[t]
\centerline{\psfig{figure=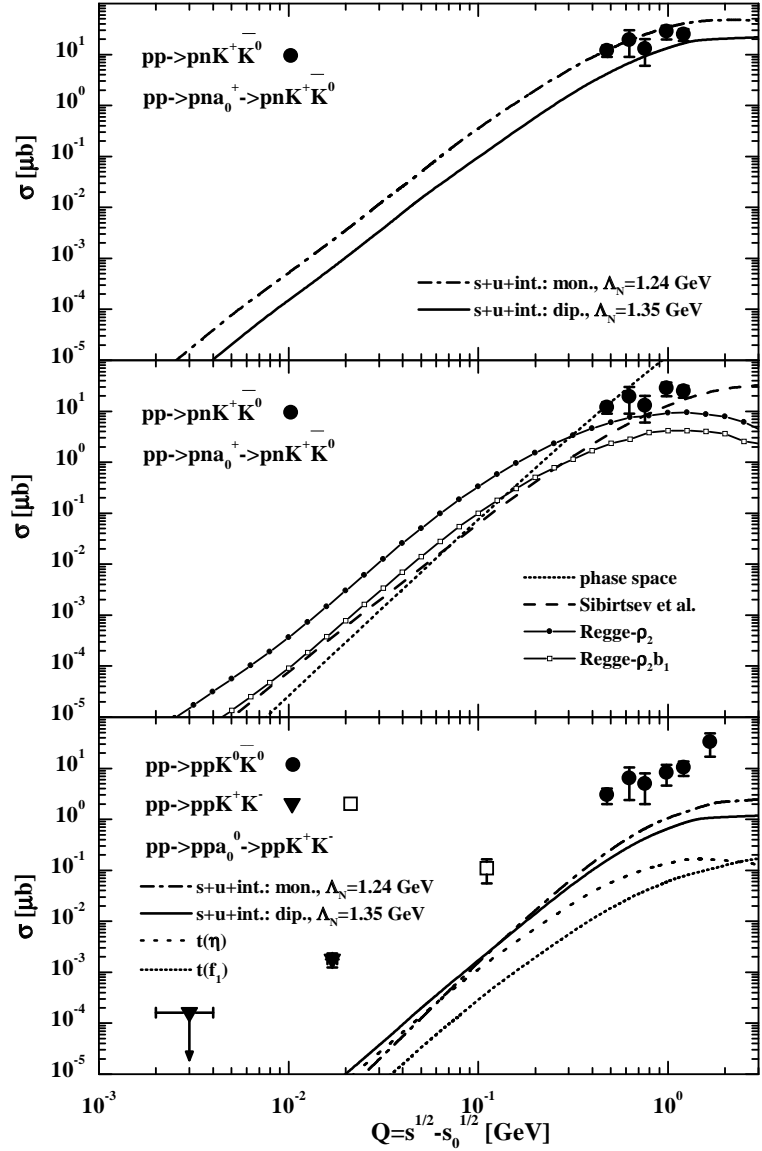,width=10cm}}
\caption{
Upper part: the calculated total cross section (within
parameter set 1 (\ref{set1})) for the reaction $pp\to pn
a_0^+ \to pn K^+ \bar K_0$ in comparison to the experimental data for
$pp \to pn K^+ \bar K_0$ (solid dots) from \protect\cite{Landolt}
as a function of $Q=\sqrt{s}-\sqrt{s_0}$.  The dot-dashed and solid
lines  correspond to the coherent sum of $s(N)$ and $u(N)$ channels
with interference ($s+u+int.$) calculated with the monopole form factor
with $\Lambda_N=1.24$~GeV and with the dipole form factor with
$\Lambda_N=1.35$~GeV, respectively.  Middle part: the solid lines with
full dots and with open squares represent the results within the
$\rho_2$ and $(\rho_2,b_1)$ Regge exchange model.  The short dashed
line shows the 4-body phase space (with constant interaction
amplitude); the dashed line is the parametrization from Sibirtsev {\it et
al.}~\protect\cite{Sibirtsev1}.  Lower part: the calculated total cross
section (within parameter set 1) for the reaction $pp\to pp a_0^0
\to pp K^+ K^-$ as a function of $Q=\sqrt{s}-\sqrt{s_0}$ in comparison
to the experimental data.  The solid dots indicate the data for $pp \to
pp K_0 \bar K_0$ from \protect\cite{Landolt}, the open square for
$pp\to pp K^+K^-$ from \protect\cite{DISTO}; the full down
triangls show the data from \protect\cite{COSY11}.
}\label{pp_kk}
\end{figure}

   In the upper part of Fig.~\ref{pp_kk} we display the calculated total
cross section (within parameter set 1 (\ref{set1})) for the
reaction $pp\to pn a_0^+ \to pn K^+ \bar K^0$ in comparison to the
experimental data for $pp \to pn K^+ \bar K^0$ (solid dots) from
\cite{Landolt} as a function of the excess energy
$Q=\sqrt{s}-\sqrt{s_0}$.  The dot-dashed and solid lines in Fig.
\ref{pp_kk} correspond to the coherent sum of $s(N)$ and $u(N)$
channels with  interference ($s+u+$int.), calculated with a monopole
form of the form factor (\ref{FN}) with $\Lambda_N=1.24$~GeV and with a
dipole form $(F_{N}(u)^2)$ with $\Lambda_N=1.35$~GeV, respectively. We
mention that the  latter (dipole) result is in better agreement with
the constraints on the near-threshold production of $a_0$ in the
reaction $\pi^+p \to K^+ \bar {K^0} p $ (see Section 3).  In the
middle part of Fig.~\ref{pp_kk} the solid lines with full dots and with
open squares present the results within the $\rho_2$ and $(\rho_2,b_1)$
Regge exchange model. The dotted line shows the 4-body phase
space (with constant interaction amplitude), while the dashed line is
the parametrization from Sibirtsev {\it et al.}  \cite{Sibirtsev1}.  We
note, that the cross sections for parameter set 2 (\ref{set2}) are
similar to set 1 (\ref{set1}) and larger by a factor $\sim 1.5$.

In the lower part of Fig.~\ref{pp_kk} we show the calculated total
cross section (within parameter set 1) for the reaction $pp\to pp
a_0^0 \to pp K^+ K^-$ as a function of $Q=\sqrt{s}-\sqrt{s_0}$ in
comparison to the experimental data. The  solid dots indicate the data
for $pp \to pp K^0 \bar K^0$ from \cite{Landolt}, the open square for
$pp\to pp K^+K^-$ is from the DISTO collaboration~\cite{DISTO}, and the
full down triangles show the data from COSY-11 \cite{COSY11}.

For the $pp \to pp a_0^0\to pp K^+K^-$ reaction (as for $pp\to  pp
a_0^0$) there is no contribution from meson Regge trajectories; $s$
and $u$ channels give similar contributions such that their
interference according to the effective OPE model (line $s+u+$int.) is
strongly destructive (cf. upper part of Fig.~\ref{pp_q}).  The
$t(f_1)$ contribution (dotted line) is practically negligible,
while the $t(\eta)$ channel (rare-dotted line) becomes important closer to the
threshold.

Thus our model gives quite small cross sections for $a_0^0$ production
in the $pp\to pp K^+K^-$ reaction which complicates its experimental
observation for this isospin channel.  The situation looks more
promising  for the $pp\to pn a_0^+ \to pn K^+\bar K^0$ reaction since
the $a_0^+$ production cross section is by an order of magnitude larger
than the $a_0^0$ one. Moreover, as has been pointed out with respect to
Fig.~\ref{pp_q}, the influence of the interference is not so strong as
for the $pp\to pp a_0^0 \to pp K^+K^-$ reaction.

Here we stress again the limited applicability of the effective
Lagrangian model (ELM) at high energies. As seen from the upper part of
Fig. \ref{pp_kk}, the ELM calculations at high energies go through the
experimental data, which is not realistic since also other channels
contribute to $K^+\bar K^0$ production in $pp$ reactions (cf. dashed
line from \cite{Sibirtsev1}). Moreover, the ELM calculations are
higher than the Regge model predictions which indicates, that the ELM
amplitudes at high energies have to be reggeized.

\subsubsection{Numerical Results for the Invariant Mass Distribution}

   As follows from the lower part of Fig.~\ref{pp_kk}, the $a_0$
contribution to the $K^+K^-$ production in the $pp\to pp K^+K^-$
reaction near the threshold is hardly seen. With increasing energy the
cross section grows up, however, even at $Q=0.111$~GeV the full cross
section with interference ($s+u+$int.) gives only a few percent
contribution to the $0.11\pm 0.009\pm 0.046\ \mu$b ``nonresonant'' cross
section (without $\phi\to K^+K^-$) from the DISTO collaboration
\cite{DISTO}.

\begin{figure}[h]
\centerline{\psfig{figure=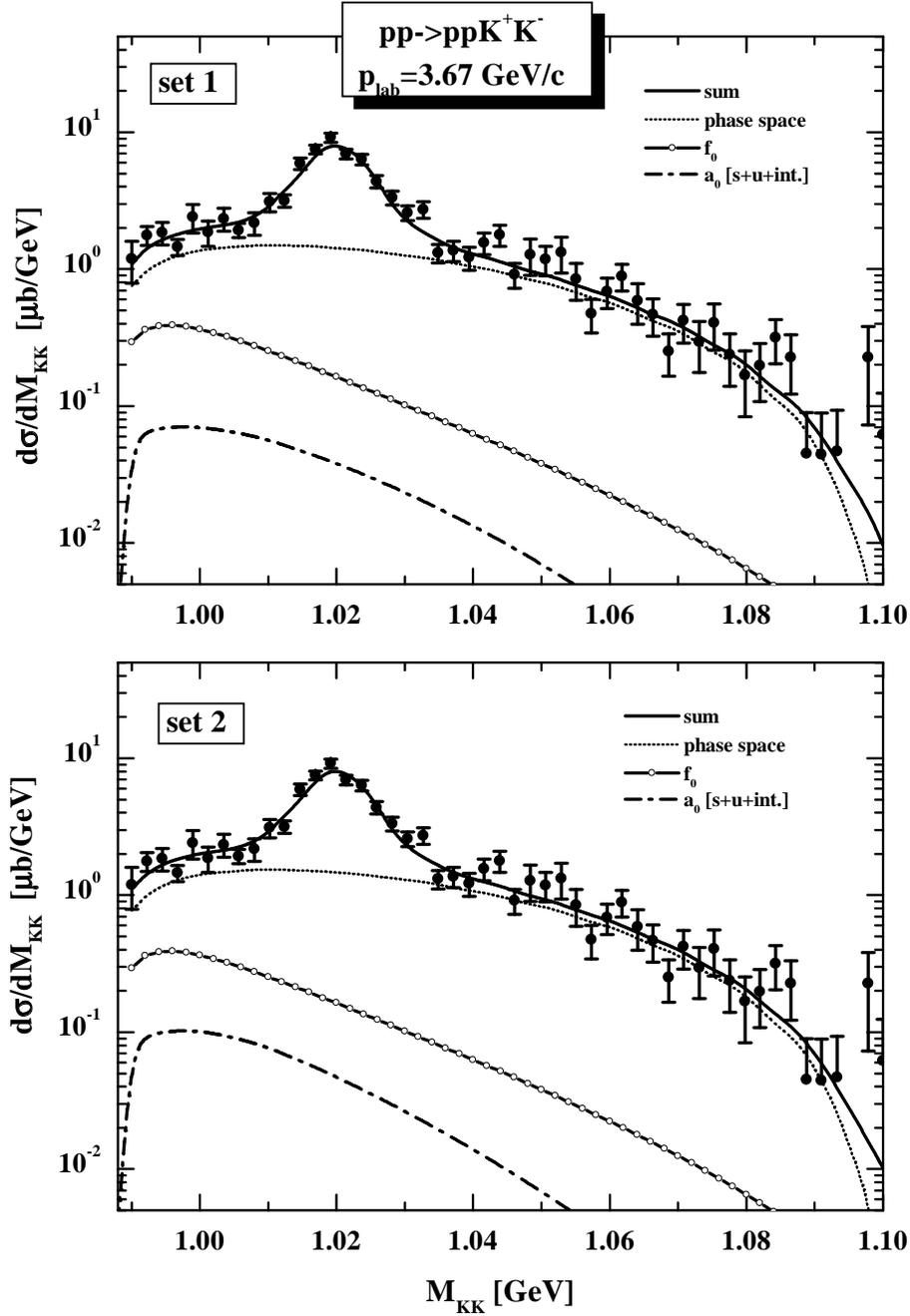,width=12cm}}
\vspace*{1cm}\caption{
The $K^+K^-$ invariant mass distribution for the $pp\to
pp K^+K^-$ reaction at $p_{lab}=3.67$~GeV$/c$.  The short dotted lines
indicate the 4-body phase space with constant interaction amplitude,
the dot-dashed lines show the coherent sum of $s(N)$ and $u(N)$
channels with interference ($s+u+int.$).  The solid lines with open
circles correspond to the $f_0$ contribution from
\protect\cite{Brat99}.  The thick solid lines show the sum of all
contributions including the decay $\phi\to K^+K^-$.  The experimental
data are taken from \protect\cite{DISTO}.
}\label{distf0a0}
\end{figure}

To clarify the situation with the relative contribution of $a_0^0$ to
the total $K^+K^-$ production in $pp$ reactions we calculate the
$K^+K^-$ invariant mass distribution for the $pp\to pp K^+K^-$ reaction
at $p_{{\rm lab}}=3.67$~GeV$/c$, which corresponds to the kinematical
conditions for the DISTO experiment \cite{DISTO}. The differential
results are presented in Fig.~\ref{distf0a0}.  The upper part  shows
the calculation within parameter set 1, whereas the lower part
corresponds to set 2.  The dot-dashed lines (lowest curves)
indicate the coherent sum of $s(N)$ and $u(N)$ channels with
interference ($s+u+$int.) for the $a_0$ contribution.  However, one has
to consider also the contribution from the $f_0$ scalar meson, i.e.
the $pp\to pp f_0\to pp K^+K^-$ reaction.  The $f_0$ production in $pp$
reactions has been studied in detail in \cite{Brat99}. Here we use
the result from  \cite{Brat99} and show in Fig.~\ref{distf0a0} the
contribution from the $f_0$ meson (calculated with parameter set $A$
from \cite{Brat99}) as the solid line with open circles ($f_0$).

We find that when adding the $f_0$ contribution to the phase-space of
nonresonant $K^+K^-$ production (the dotted lines in
Fig.~\ref{distf0a0}) and the contribution from $\phi$ decays (resonance
peak around 1.02 GeV), the sum (solid) lines almost perfectly describe
the DISTO data.  This means that there is no visible signal for an
$a_0^0$ contribution in the DISTO data according to our calculations
while  the $f_0$ meson gives some contribution to the $K^+K^-$
invariant mass distribution at low invariant masses $M$, that is $\sim
12\%$ of the total ``nonresonant'' cross section from the DISTO
collaboration \cite{DISTO}. Thus the reaction $pp \to pn K^+ \bar K^0$
is more promising for $a_0$ measurements as has been pointed  above.

\subsubsection{Nonresonant Background}

\begin{figure}[h]
\centerline{\psfig{file=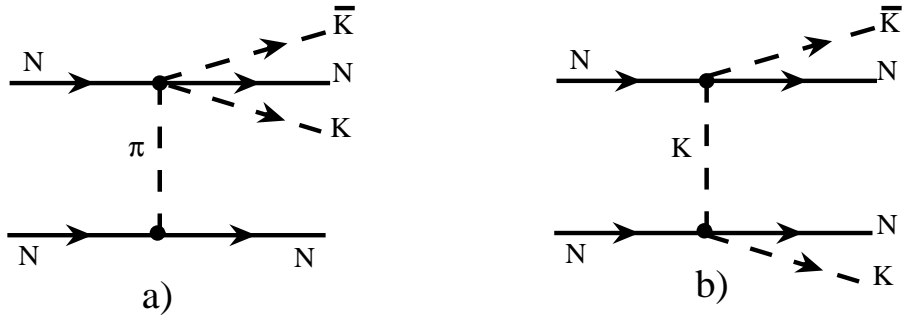,width=12cm}}
\vspace*{1cm}\caption{
The diagrams a)-b) describing  different mechanisms of
nonresonant $K\bar K$ production in the reaction $NN\to NN K \bar K$.
}\label{fig10r}
\end{figure}

   Following \cite{Sibirtsev1} we consider two mechanisms of
nonresonant $K \bar K$ production, related to pion and kaon exchanges,
which are described by the diagrams a) and b) in Fig.~\ref{fig10r}.
The pion exchange amplitude can be calculated using the results of
Section 3.  As concerning the kaon exchange mechanism, the amplitude of
the reaction  $NN \to NN a_0 \to NN K \bar K$ can be written as
\begin{eqnarray}
&&\hspace*{-4mm}\mathrm{M_{K-exchange}}(p_a,p_b;p_c,p_d,k_1,k_2) =
\frac{F_{K}^2(q^2)}{q^2-m_K^2} \times \nonumber \\
&\times&
\bar u(p_c) \ A_{KN \to KN}(p_c,k_1;p_a,q)\ u(p_a) \times \nonumber \\
&\times&
\bar u(p_d) \ A_{\bar{K}N \to \bar{K}N}(p_d,k_2;p_b,q) \ u(p_b)
\label{KKNN}
\end{eqnarray}
with permutations of nucleons in the initial and final states.
Here $p_a, p_b$ and $p_c, p_d$ are the four momenta of the initial and
final nucleons, respectively; $k_1$ and $k_2$ are the momenta of the
final kaons; $q$ is the momentum of the virtual kaon; $F_{K}(q^2)$ is
the kaon form factor which we take in the monopole form with the cut-off
parameter $\Lambda$ =1.2 GeV.

The antikaon--nucleon amplitude $A_{\bar{K}N \to \bar{K}N}$ has been
taken from \cite{Martin} explicitly.  Since near threshold the
$KN\to KN$ cross section depends mainly on the normalization of the
amplitude, but not on its spin dependence, we adopt the simplest
approximation that the amplitude  $A_{KN \to KN}$ is a Lorentz scalar.
This allows us to connect the $A_{KN \to KN}$ amplitude (squared)
by simple kinematical factor to the $KN\to KN$ cross section, where the
parametrization for the elastic $K^+p\to K^+p$ cross section has been
taken from \cite{Gugnon} and the $K^0p\to K^+n$ cross section has
been parametrized according to the existing data \cite{Landolt,Armitage77}.

\begin{figure}[h]
\centerline{\psfig{figure=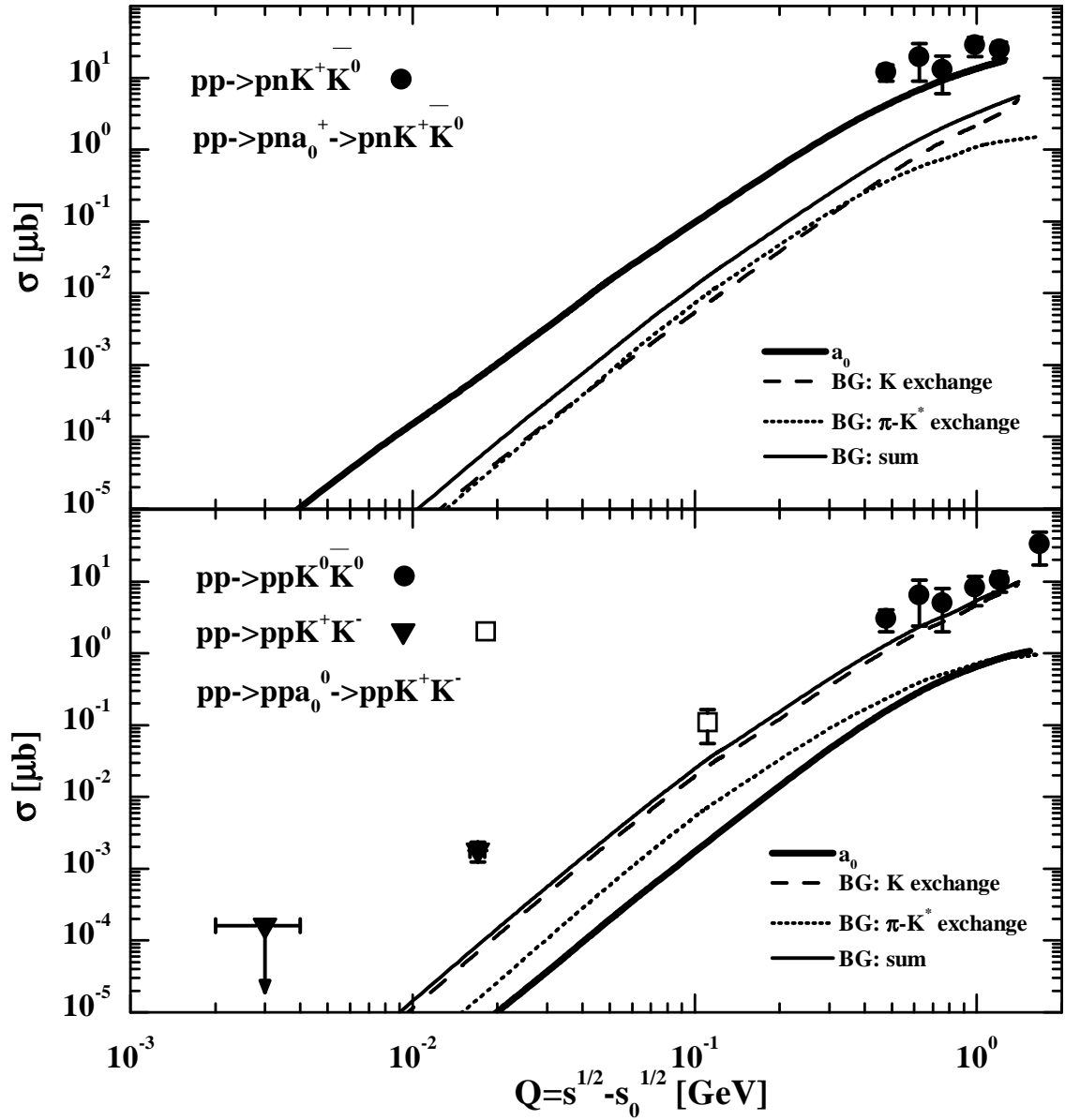,width=15cm}}
\vspace*{1cm}\caption{
Comparison of the $a_0$-resonance contribution (bold solid
curves) and nonresonant background (thin solid curves) in the reactions
$pp\to pnK^+ \bar{K^0}$ (upper part) and $pp \to pp K^+ K^-$ (lower part).
}\label{fig11r}
\end{figure}

The results of our calculations are shown in Fig.~\ref{fig11r} in
comparison to the experimental data. The contribution of the pion
exchange mechanism (which we denoted as ``BG:$\pi-K^* ~exchange$'') is shown
by the dotted curves.  The dashed lines in the upper and lower parts
describe the $K$-exchange mechanism. The thin solid lines show the
total background, which in our model is the sum of pion and kaon
exchange contribution.  This background can be compared with the $a_0$
production cross section shown by the bold solid lines. In the case of
the reaction $pp \to pnK^+ \bar{K^0}$ (upper part) the $a_0$
production cross section is much larger than the background, while in
the case of the reaction $pp \to ppK^+ K^-$ (lower part) the
$a_0(980)$ resonance contribution (bold solid line) appears to be
much smaller than the nonresonant background.  We mention that the
disagreement with the DISTO ($Q \simeq 100$ MeV) and COSY--11 ($Q \simeq
17$ MeV) data should be related to the $K^-pp$ final state interaction,
which is known to be strong.

\subsubsection{Concluding Remarks on $a_0$ Production in $pN$ Reactions}

   In this Section we have estimated the cross sections of  $a_0$ production
in the reactions $pp\to pp a_0^0$ and $pp\to pn a_0^+$
near threshold and at medium energies.  Using an
effective Lagrangian approach with one-pion exchange we have analyzed
different contributions to the cross section corresponding to
$t$-channel diagrams with $\eta(550)$- and $f_1(1285)$-meson exchanges
as well as $s$- and $u$-channel graphs with an intermediate nucleon.
We additionally have considered  the $t$-channel Reggeon exchange
mechanism with parameters normalized to the Brookhaven data for
$\pi^-p\to a_0^-p$ at 18 GeV$/c$ \cite{Brookhaven}.  These results have been
used to calculate the contribution of $a_0$ mesons to the cross
sections of the reactions  $pp\to pn K^+\bar K^0$ and $pp\to pp
K^+K^-$.  Due to unfavorable isospin Clebsh--Gordan coefficients as
well as rather strong destructive interference of the $s$- and
$u$-channel contributions  our model gives quite small cross sections
for $a_0^0$ production in the $pp\to pp K^+K^-$ reaction.  However, the
$a_0^+$ production cross section in  the $pp\to pn a_0^+ \to pn K^+\bar
K^0$ reaction should be larger by about an order of magnitude.
Therefore the experimental observation of $a_0^+$ in the reaction
$pp\to pn K^+\bar K^0$ is much more promising than the observation of
$a_0^0$ in the reaction $pp\to pp K^+K^-$.  We note in passing that the
$\pi\eta$ decay channel is experimentally more challenging since, due
to the larger nonresonant background \cite{Mueler01}, the
identification of the $\eta$-meson (via its decay into photons) in a
neutral-particle detector is required.

We have also analyzed invariant mass distributions of the $K \bar K$
system  in the reaction $pp\to pN a_0 \to pN K \bar K$ at different
excess energies $Q$ not far from threshold.  Our analysis of the DISTO
data on the reaction $pp \to pp K^+K^-$ at 3.67 GeV$/c$ has shown that
the $a_0^0$ meson is practically not seen in $d\sigma /dM$ at low
invariant masses, however, the $f_0$ meson gives some visible
contribution.  In this respect the possibility to measure the $a_0^+$

meson in $d\sigma /dM$ for the reaction $pp\to pn K^+\bar K^0$ (or $\to
d K^+\bar K^0$) looks much more promising not only due to a much larger
contribution for the  $a_0^+$, but also due to the absence of the $f_0$
meson in this channel. It is also very important that the nonresonant
background  is expected to be much smaller than the $a_0$ signal in the
$pp\to pn K^+\bar K^0$ reaction.

Experimental data on $a_0$ production in $NN$ collisions are
practically absent (except of the $a_0$ observation in the reaction
$pp\to dX$ \cite{BNL73}).  Such measurements might give new information
on the $a_0$ structure. According to Atkinson {\it et al.}
\cite{Atkinson} a relatively strong production of the $a_0$ (the same
as for the $b_1(1235)$) in non-diffractive reactions can be considered
as evidence for a $q \bar q$ state rather than a $qq \bar q \bar q$
state. For example, the cross section of $a_0$ production in $\gamma p$
reactions at 25--50 GeV is about 1/6 of the cross sections for $\rho$
and $ \omega$ production.  Similar ratios are found  in the two-body
reaction $pp \to d X$ at 3.8--6.3 GeV$/c$ where $\sigma (pp \to d
a_0^+) =(1/4 - 1/6)\sigma (pp \to d \rho^+)$.

In our case we can compare $a_0$ and $\omega$ production.  Our model
predicts  $\sigma (pp \to pn a_0^+) = 30-70\  \mu$b at $Q \simeq 1$
GeV which can be compared with $\sigma (pp \to
pp\omega) \simeq  100-200 \ \mu$b at the same $Q$.  If such a large
cross section could be detected experimentally this would be a serious
argument in favor of the $q \bar q$ model for the  $a_0$.

To distinguish between the threshold cusp scenario and a resonance
model one can exploit different analytical properties of the $a_0$
production amplitudes. In case of a genuine resonance the amplitude of
$\eta \pi$ and $K \bar K$ production through the $a_0$ has a pole and
satisfies the factorization property.  This implies that the shapes of
the invariant mass distributions in the $\eta \pi$ and $K \bar K$
channels should not depend on the specific reaction in which the $a_0$
resonance is produced (for $Q \geq \Gamma_{\rm tot}$). On the other hand,
for the threshold cusp scenario the $a_0$ bump is produced through the
$\pi \eta$ final state interaction.  The corresponding amplitude has a
square root singularity and in general can not be factorized (see, e.g.,
\cite{BaruFSI} were the factorization property was disproven for
$pp$ FSI in the reaction $pp \to pp M$).  This implies that for a
threshold bump the invariant mass distributions in the $\eta \pi$ and
$K \bar K$ channels are expected to be different for different
reactions and will depend on kinematical conditions (i.e., momentum transfer)
even at the same value of excess energy, e.g., $Q\simeq 1$ GeV.

\section{$a_0(980)$-$f_0(980)$ Mixing and Isospin Violation in the
Reactions $pN \rightarrow d a_0$, $pd \rightarrow \mathrm{^3He/^3H}\,
a_0$ and $ dd \rightarrow \mathrm{^4He}\, a_0$}

\subsection{Hints for $a_0(980)$--$f_0(980)$ Mixing}

   As it was suggested long ago in \cite{Achasov} the dynamical
interaction of the $a_0(980)$- and $f_0(980)$-mesons with states
close to the $K \bar K$ threshold may give rise to a significant
$a_0(980)$--$f_0(980)$ mixing. Different aspects of this mixing and
the underlying  dymanics as well as the possibilities to measure
this effect have been  discussed in
\cite{Jan},\cite{Achasov2}--\cite{Grishina2001}, \cite{Kudryavtsev}.
Furthermore, it has been  suggested  by Close and Kirk~\cite{Close2000}
that the new data from
the WA102 collaboration at CERN~\cite{WA102} on the central
production of $f_0$ and $a_0$ in the reaction $pp\rightarrow p_s X
p_f$ provide evidence for a significant $f_0$--$a_0$ mixing
intensity  as large as $|\xi|^2=8\pm 3$\%.
In this Section we will discuss possible experimental tests of this
mixing in the reactions
$$pp\rightarrow da_0^+~~(a),~~pn \rightarrow da_0^0~~(b),$$
$$pd\rightarrow \mathrm{^3H}\, a_0^+~~(c),~~pd \rightarrow
\mathrm{^3He}\, a_0^0~~(d)$$
and
$$dd\rightarrow \mathrm{^4He}\, a_0^0~~ (e)$$
near the corresponding thresholds. We recall that the $a_0$-meson can
decay to $\pi \eta$ or $K \bar K$. Here we only consider  the
dominant $\pi \eta$ decay mode.  Note that the isospin violating
anisotropy in the reaction $pn \rightarrow da_0^0$ due to the
$a_0(980)$--$f_0(980)$ mixing is very similar to that which might arise in
the reaction $pn \rightarrow d \pi^0$ because of the $\pi^0$--$\eta$
mixing (see \cite{Tippens}). Recently measurements of the
charge-symmetry breaking in the reactions $\pi^+ d \rightarrow pp \eta$
and  $\pi^- d \rightarrow nn \eta$ near the $\eta$ production threshold
were performed at BNL \cite{Tippens}. A similar experiment, comparing the
reactions $p d \rightarrow \mathrm{^3 He} \pi^0$ and $p d \rightarrow
\mathrm{^3 H} \pi^+$ near the $\eta$ production threshold, is now in
preparation at COSY (J\"ulich) ( see, e.g., \cite{Magiera}).

\subsection{Reactions $pp\rightarrow da_0^+$ and
                       $pn \rightarrow da_0^0$ }

\subsubsection{Phenomenology of Isospin Violation}

   In the reactions (a) and (b) the final $da_0$ system has isospin $I_f=1$,
for $l_f=0$ ($S$-wave production close to threshold) it has spin--parity
$J^P_f=1^+$. The initial $NN$ system cannot be in
the state $I_i=1,~J^P_i=1^+$ due to the Pauli principle. Therefore,
near threshold the $da_0$ system should be dominantly produced in
$P$-wave with quantum numbers $J_f^P=0^-,~1^-$ or $2^-$. The states with
$J_i^P=0^-,~1^-$ or $2^-$ can be formed by an $NN$ system with
spin $S_i=1$ and $l_i=1$ and 3. At the beginning for qualitative discussion
we neglect the contribution of the
higher partial wave ($l_i=3$)\footnote{See, e.g., phenomenological analysis in \cite{Kudryavtsev2002} where this partial wave was also taken into account.}. In this case we can write the amplitude of reaction (a) in the following form\\
\begin{eqnarray}
  &&T(pn \rightarrow d~a_0^+)= \nonumber \\
  &&=\alpha^{+}\ {\bf {p \cdot S}}\ {\bf {k \cdot e}}{}^* +\beta^{+}\
{\bf {p \cdot k}}\ {\bf {S \cdot e}}{}^* +\gamma^{+}\ {\bf {S \cdot k}}\
{\bf {p \cdot e}}{}^*,
\end{eqnarray}
where ${\bf S}=\phi_N^T \sigma_2 \
\mbox{\boldmath$\sigma$}\phi_N$ is the spin operator of the
initial $NN$ system; $\bf{p}$ and $\bf{k}$ are the initial and
final c.m. momenta; $\bf e$ is the deuteron
polarization vector; $\alpha^+$, $\beta^+$, $\gamma^+$ are three
independent scalar amplitudes which can be considered as constants
near threshold (at $k \rightarrow 0$).

Due to the mixing, the $a_0^0$ may also be produced via the $f_0$. In
this case the $a_0^0 d$ system will be in $S$-wave and the
amplitude of reaction (b) can be written as:
\begin{eqnarray}
  &&T(pn \rightarrow d~a_0^0)= \nonumber \\
  &&=\alpha^{0}\ {\bf {p \cdot S}}\ {\bf {k \cdot e}}{}^* +\beta^{0}\
{\bf {p \cdot k}}\ {\bf {S \cdot e}}{}^* +\gamma^{0}\ {\bf {S \cdot k}}\
{\bf {p \cdot e}}{}^* + \xi F \ \bf {S \cdot e}{}^*,
\end{eqnarray}
where $\xi$ is the mixing parameter and $F$ is the $f_0$ production
amplitude. In the limit $k \rightarrow 0$, $F$ is again a constant.
The scalar amplitudes $\alpha$, $\beta$, $\gamma$ for reactions (a)
and (b) are related to each other by a relative factor of $\sqrt{2}$
as: $\alpha^{+}=\sqrt{2} \alpha^0$, $\beta^{+}=\sqrt{2} \beta^0$,
$\gamma^{+}=\sqrt{2} \gamma^0$.

The differential cross sections for reactions (a) and (b) have the
form (up to terms linear in $\xi$)
\begin{eqnarray}
  &\displaystyle \frac{{\mathrm{d}}\sigma(pp\rightarrow
    d~a_0^+)}{{\mathrm{d}}\Omega}=&
  2\ \frac{k}{p}\left(C_0+C_2 \cos^2 \Theta \right) \label{pp},\\
  &\displaystyle \frac{{\mathrm{d}}\sigma(pn\rightarrow d ~a_0^0)}
  {{\mathrm{d}}\Omega}=& \frac{k}{p}\left(C_0+ C_2 \cos^2 \Theta
  \right.+\left. C_1 \cos \Theta \right)\ ,\label{pn}
\end{eqnarray}
where
\begin{eqnarray}
\label{coeff} &&C_0=\frac{1}{2}\ p^2 k^2
\left[|\alpha^0|^2+|\gamma^0|^2 \right] ,~ C_1=p\ k
\left[\mathrm{Re} ((\xi F)^{*}(\alpha^0 +3\,
\beta^0+\gamma^0))\right] \, \nonumber \\ &&C_2=\frac{1}{2}\ p^2
k^2\left[ 3\, |\beta^0|^2\right. \left. +2\, \mathrm{Re} (\alpha^0
\beta^{0\, *}+\alpha^0 \gamma^{0\, *}+ \beta^0 \gamma^{0\,
*})\right] \ .
\end{eqnarray}
Similarly, the differential cross section of the reaction $pn
\rightarrow d f_0$ can be written as
\begin{equation}
  \frac{{\mathrm{d}}\sigma( pn \rightarrow d f_0)}
  {{\mathrm{d}}\Omega}= \frac{3\, k}{2\, p}\ |F|^2\ . \nonumber
\end{equation}
The mixing effect --- described by the term $C_1\cos \Theta$ in
Eq. (\ref{pn}) --- then leads to an isospin violation in the ratio
$R_{ba}$ of the differential cross sections for reactions (b)
and (a),
\begin{equation}
R_{ba}=\frac12+\frac{C_1 \cos\Theta}{C_0+C_2\cos ^2\Theta} ,
\nonumber
\end{equation}
and to the forward--backward asymmetry for reaction (b):
\begin{equation}
A_b(\Theta)=\frac{\sigma_b(\Theta)-\sigma_b(\pi- \Theta)}
{\sigma_b(\Theta)+\sigma_b(\pi-\Theta)}=
\frac{C_1 \cos\Theta}{C_0+C_2\cos^2\Theta}\nonumber \ .
\label{asym}
\end{equation}

The latter effect has been already discussed in \cite{Kudryavtsev}
where it was argued that the asymmetry $A_b(\Theta=0)$ can reach (5--
10)\% at an energy excess of $Q=(5- 10)$~MeV. However, if we adopt a
mixing parameter $|\xi|^2=(8\pm 3)$\%, as it follows from the WA102
data, we can expect a much larger asymmetry. We note explicitly, that
the coefficient $C_1$ in (\ref{coeff}) depends not only on the
magnitude of the mixing parameter $\xi$, but also on the relative
phases with respect to the amplitudes of $f_0$ and $a_0$ production,
which are unknown so far. This uncertainty has to be kept in mind for
the following discussion.

If $a_0$ and $f_0$ were very narrow particles, then near threshold the
differential cross section (\ref{pp}), dominated  by the $P$-wave, would
be proportional to $k^3$ or $Q^{3/2}$, where $Q$ is the c.m. energy
excess. Due to $S$-wave dominance in the reaction $pn\rightarrow d f_0$
one would expect that the cross section scales like $\sim k $ or $\sim
\sqrt{Q}$.  In this limit the $a_0$--$f_0$ mixing leads to an enhancement
of the asymmetry $A_b(\Theta)$ as $1/k$ near threshold. In reality,
however, both $a_0$ and $f_0$ have widths of about 40$ - $100 MeV.
Therefore, at fixed initial momentum their production cross section
should be averaged over the corresponding mass distributions.  This
will essentially change the threshold behavior of the cross sections.
Another complication is that broad resonances are usually accompanied
by background lying underneath the resonance signals. These problems
will be discussed below in the following Subsections.

\subsubsection{Model Calculations}

In order to estimate isospin-violation effects in the differential
cross-section ratio $R_{ba}$ and in the forward--backward asymmetry
$A_b$ we use the two-step model (TSM), which was successfully applied
earlier to the description of $\eta$-, $\eta^{\prime}$-, $\omega$- and
$\phi$-meson production in the reaction $pN \rightarrow d X$ in
\cite{Grishina1,Grishina2}. Recently, this model has been also used for
an analysis of the reaction $pp \rightarrow d a_0^+$ \cite{Grishina3}.

\begin{figure}[h]
\centerline{\psfig{file=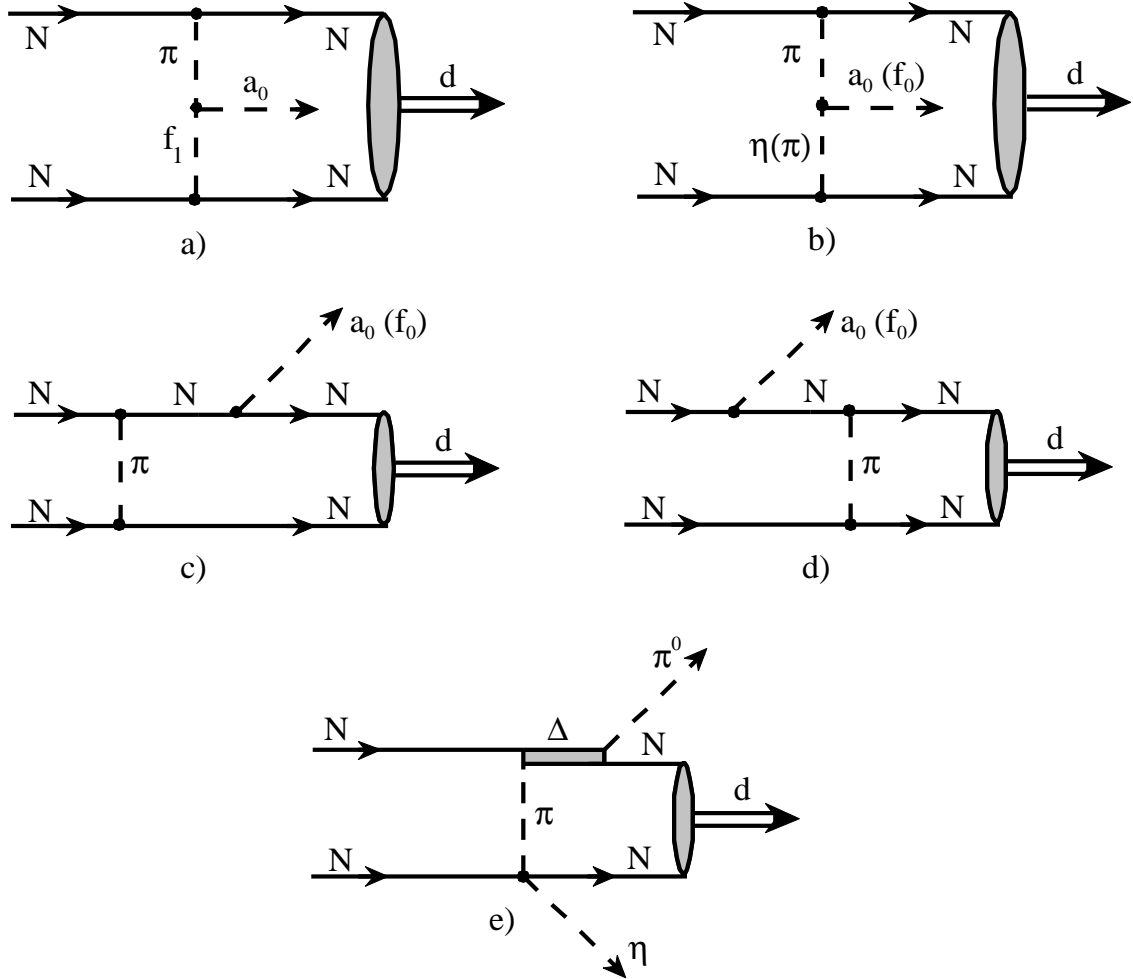,width=15cm}}
\vspace*{1cm}\caption{
Diagrams a)-d) describing different mechanisms of $a_0$
and $f_0$-meson production in the reaction $NN\to da_0(f_0)$ within
the framework of the two-step model (TSM). The nonresonant $\pi\eta$
production is described by the diagram e).
}\label{fig:tsm}
\end{figure}

The diagrams in Fig.~\ref{fig:tsm} describe the different mechanisms of
$a_0$- and $f_0$-meson production in the reaction $NN\to da_0/f_0$
within the framework of the TSM. In the case of $a_0$ production the
amplitude of the subprocess $\pi N \to a_0 N$ contains three different
contributions: i) the $f_1(1285)$-meson exchange (Fig.~\ref{fig:tsm}
{\it a}); ii) the $\eta$-meson exchange (Fig.~\ref{fig:tsm} {\it b});
iii) $s$- and $u$-channel nucleon exchanges (Fig.~\ref{fig:tsm} {\it c}
and \ref{fig:tsm} {\it d}).  As it was shown in \cite{Grishina3} the
main contribution to the cross section for the reaction $pp \rightarrow
d a_0^+$ stems from the $u$-channel nucleon exchange (i.e., from the
diagram of Fig.~\ref{fig:tsm} {\it d}) and all other contributions can
be neglected. In order to preserve the correct structure of the
amplitude under permutations of the initial nucleons (which is
antisymmetric for the isovector state and symmetric for the isoscalar
state) the amplitudes of $a_0$ and $f_0$ production can be written as
the following combinations of the $t$- and $u$-channel contributions:
\begin{eqnarray}
&&T_{pn\to da_0^0}(s,t,u) = A_{pn\to da_0^0}(s,t)-A_{pn \to
 da_0^0}(s,u),\nonumber\\ &&T_{pn\to df_0}(s,t,u) = A_{pn \to
df_0}(s,t)+A_{pn\to df_0}(s,u),\label{Atu}\end{eqnarray}
where $s=(p_1+p_2)^2$, $t=(p_3-p_1)^2$, $u=(p_3-p_2)^2$ and $p_1$,
$p_2$, $p_3$, and $p_4$ are the 4-momenta of the initial protons, meson
$M$ and the deuteron, respectively. The structure of the amplitudes
(\ref{Atu}) guarantees that the $S$-wave part vanishes in the case of
direct $a_0$ production since it is forbidden by angular momentum
conservation and the Pauli principle.  Also higher partial waves are
included in (\ref{Atu})(in contrast to the simplified discussion in Section 5.1).

In the case of $f_0$ production the amplitude of the subprocess $\pi N
\to f_0 N$ contains two different contributions: i) the $\pi$-meson
exchange (Fig.~\ref{fig:tsm} b); ii) $s$- and $u$-channel nucleon
exchanges (Fig.~\ref{fig:tsm} {\it c}) and \ref{fig:tsm} {\it d}). Our
analysis has shown that similarly to the case of $a_0$ production the
main contribution to the cross section of the reaction $pn \rightarrow
d f_0$ is due to the $u$-channel nucleon exchange (i.e., from the
diagram of Fig.~\ref{fig:tsm} {\it d}); the contribution of the
combined $\pi\pi$ exchange (Fig.~\ref{fig:tsm} {\it b}) as well as the
$s$-channel nucleon exchange  can be neglected.  In this case we get
for the ratio of the squared amplitudes
\begin{equation}
\frac{{|A_{pn\to df_0}}(s,t)|^2}{|A_{pn\to da_0}(s,t)|^2}=
\frac{{|A_{pn\to df_0}}(s,u)|^2}{|A_{pn\to da_0}(s,u)|^2}=
\frac{|g_{f_0NN}|^2}{|g_{a_0NN}|^2} .
\label{f0a0}
\end{equation}
If we take $g_{a_0NN}$ = 3.7 (see, e.g., \cite{Holinde}) and
$g_{f_0NN}$ =8.5  \cite{Bonnf1}, then
we find for the ratio of the amplitudes $R(f_0/a_0) =
g_{f_0NN}/g_{a_0NN} =2.3$. Note, however, that Mull and Holinde \cite{Bonnf1}
give a different value for the ratio of the coupling constants
$R(f_0/a_0)=1.46$
which is lower by about 37 \%. In the following we use
$R(f_0/a_0)=$1.46--2.3.

\begin{figure}[h]
\centerline{\psfig{file=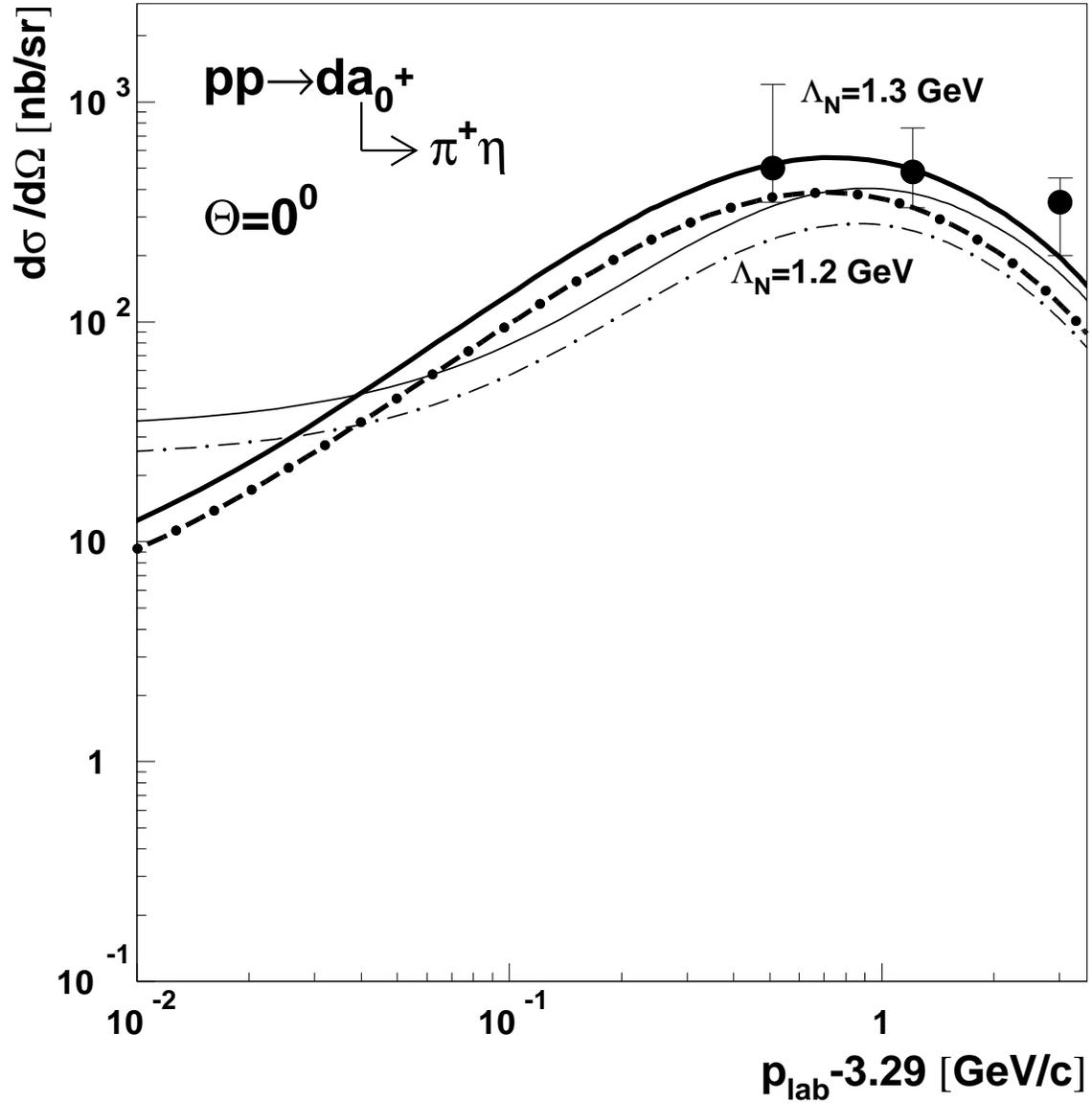,width=15cm}}
\vspace*{1cm}\caption{
Forward differential cross section of the reaction
$pp\rightarrow d a_0^+$ as a function of $(p_{\rm{lab}} - 3.29)$
GeV$/c$. The full dots are the experimental data from
\protect\cite{BNL73} while the bold dash-dotted and solid lines
describe the results of the TSM for $\Lambda_N = 1.2$ and 1.3 GeV,
respectively.  The thin dash-dotted and solid lines are calculated
using the Flatt\'e mass distribution for the $a_0$ meson with a cut $M
\geq$ 0.85 GeV.
}\label{fig:dsacosy}
\end{figure}

The forward differential cross section for reaction (a) as a function
of the proton beam momentum is presented in Fig.~\ref{fig:dsacosy}.
The bold dash-dotted and solid lines  (taken from \cite{Grishina3}
and calculated for the zero width limit) describe the results of the
TSM for different values of the nucleon cut-off parameter,
$\Lambda_N=1.2$ and 1.3 GeV, respectively.

In order to take into account the finite width of $a_0$ we use  a
Flatt\'e mass distribution with the same parameters as in
\cite{Brat01}: the $K$-matrix pole at 999 MeV, $\Gamma_{a_0 \to \pi
\eta}=$ 70 MeV, $\Gamma(K \bar K)/ \Gamma(\pi \eta)$ = 0.23 (see also
\cite{PDG} and references therein).  The thin dash-dotted and solid
lines in Fig.~\ref{fig:dsacosy} are calculated within TSM using this
mass distribution with the cut $M(\pi^+ \eta) \geq 0.85$ GeV and
$\Lambda _N=1.2$ and 1.3 GeV, respectively.  The corresponding  $\pi^0
\eta$ invariant mass distribution for the reaction $pn \rightarrow
da_0^0 \rightarrow d \pi^0 \eta$ at 3.4 GeV$/c$ is shown in
Fig.~\ref{fig:MM} by the dashed line.

In the case of the $f_0$ meson, where ${\rm Br}(K \bar K)$ is not yet fixed
\cite{PDG}, we use the Breit-Wigner mass distribution with
$m_R=980$ MeV and $\Gamma_R \simeq \Gamma_{f_0 \to \pi \pi}=$ 70 MeV.

The calculated total cross sections for the reactions $pn \to da_0$ and
$pn \to df_0$ (as a function of $T_{\mathrm{lab}}$ for $\Lambda_N$=1.2
GeV ) are shown in Fig.~\ref{fig:siga0f0}. The solid and dashed lines
describe the calculations with zero and finite widths, respectively. In
the case of $f_0$ production in the $\pi \pi$ mode we take the same cut
in the invariant mass of the $\pi \pi $ system, $M_{\pi \pi} \geq 0.85$
GeV. The lines denoted by 1 and 2 are obtained for $R(f_0/a_0)=1.46$
and 2.3. Comparing the solid and dashed lines we see that near the
threshold the finite width corrections to the cross sections are quite
important. The most important changes are introduced to the energy
behavior of the $a_0$ production cross section. (Compare also bold and
thin lines in Fig. \ref{fig:dsacosy}).

In principle, mixing can modify the mass spectrum of the $a_0$ and
$f_0$. However, in this case the effect is expected to be less
spectacular than for the $\rho$--$\omega$ case where the widths of $\rho$
and $\omega$ are very different (see, e.g., the discussion in
\cite{Tippens} and references therein). Nevertheless, the
modification of the $a_0^0$ spectral function due to $a_0$--$f_0$ mixing
can be measured comparing the invariant mass distributions of $a_0^0$
with that of $a_0^+$.  According to our analysis, a much cleaner signal
for isospin violation can be obtained from the measurement of the
forward--backward asymmetry in the reaction $pn \to d a_0^0 \to d \pi^0
\eta$ for the integrated strength of the $a_0$.  That is why for all
calculations on isospin violation effects below, the strengths of $f_0$
and $a_0$ are integrated over the invariant masses  in the interval
0.85$ - $1.02 GeV.

The magnitude of the isospin violation effects is shown in
Fig.~\ref{fig:dsa0f0}, where we present the differential cross section
of the reaction $pn \to d a_0^0$ at $T_p=2.6$ GeV as a function of
$\Theta_{\mathrm{c.m.}}$ for different values of the mixing intensity
$|\xi|^2$: 0.05 and 0.11. For reference, the solid line shows the
case of isospin conservation, i.e.,  $|\xi|^2=0$.  The dash-dotted
curves include the mixing effect. Note that all curves in
Fig.~\ref{fig:dsa0f0} were calculated assuming maximal interference of
the amplitudes describing the direct $a_0$ production and its
production through $f_0$.  The maximal values of the differential cross
section may also occur at $\Theta_{\mathrm{c.m.}}=0^{\circ}$ depending
on the sign of the coefficient $C_1$ in Eq.~(36).

It follows from Fig. \ref{fig:dsa0f0} in either case that  the
isospin-violation parameter $A_b(\Theta)$ for
$\Theta_{\mathrm{c.m.}}=180^o$ may be quite large, i.e.,
\begin{equation}
A_b(180^\circ)= 0.86-0.96~~ \mathrm{or}~~ 0.9-0.98
\label{asymmax}
\end{equation}
for  $R(f_0/a_0) $= 1.46 or 2.3, respectively.
Note that the asymmetry depends rather weakly on $R(f_0/a_0)$.
It might be more sensitive to the relative phase of
$a_0$ and $f_0$ contributions.

\subsubsection{Background}

\begin{figure}[h]
\centerline{\psfig{file=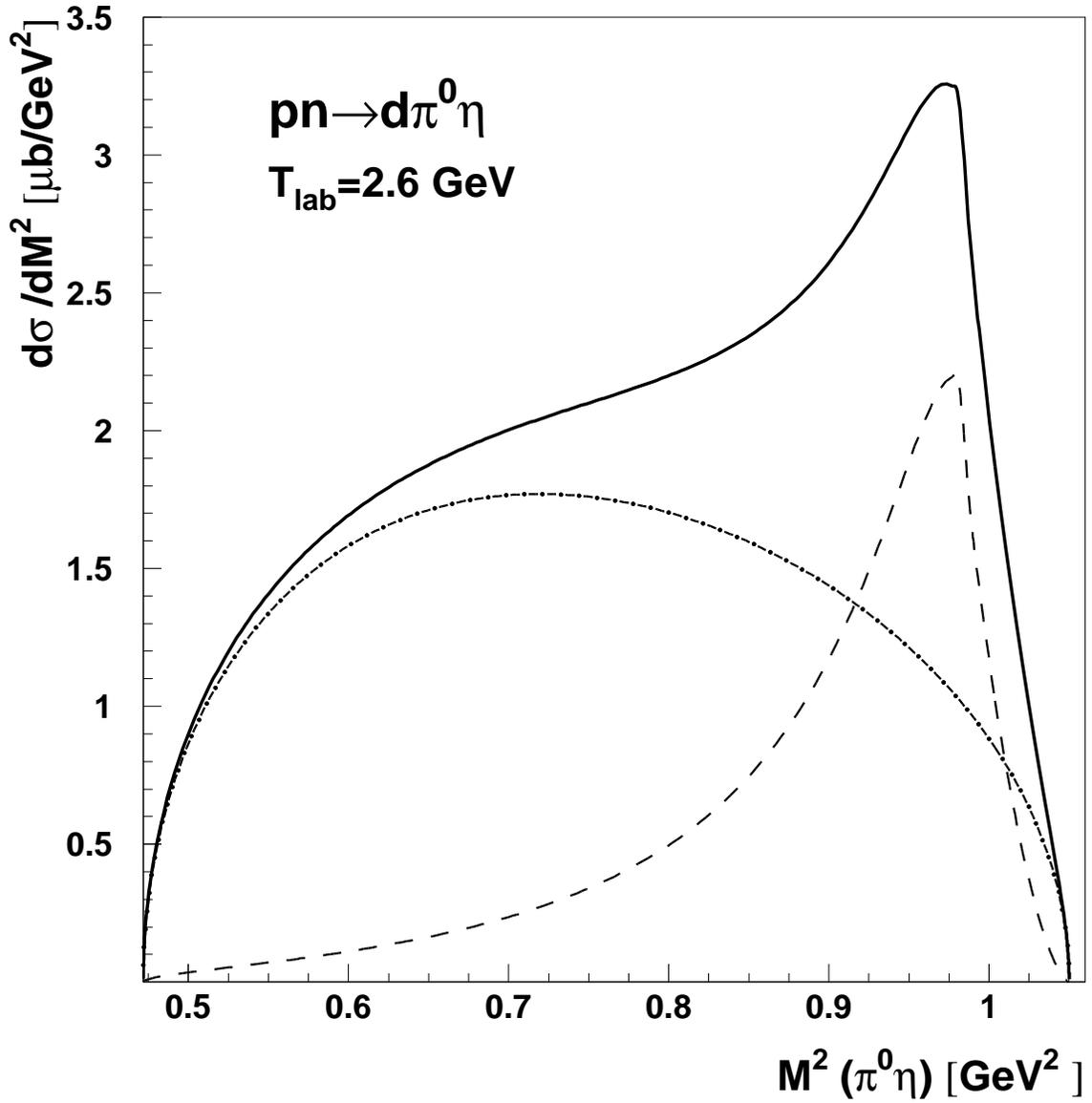,width=15cm}}
\vspace*{1cm}\caption{
$\pi^0 \eta$ invariant mass distribution for the reaction $pn
\rightarrow d \pi^0 \eta$ at 3.4 GeV$/c$. The dashed and dash-dotted
lines describe the $a_0$ resonance contribution and nonresonance
background, respectively. The solid line is the sum of both
contributions.
}\label{fig:MM}
\end{figure}

\begin{figure}[h]
\centerline{\psfig{file=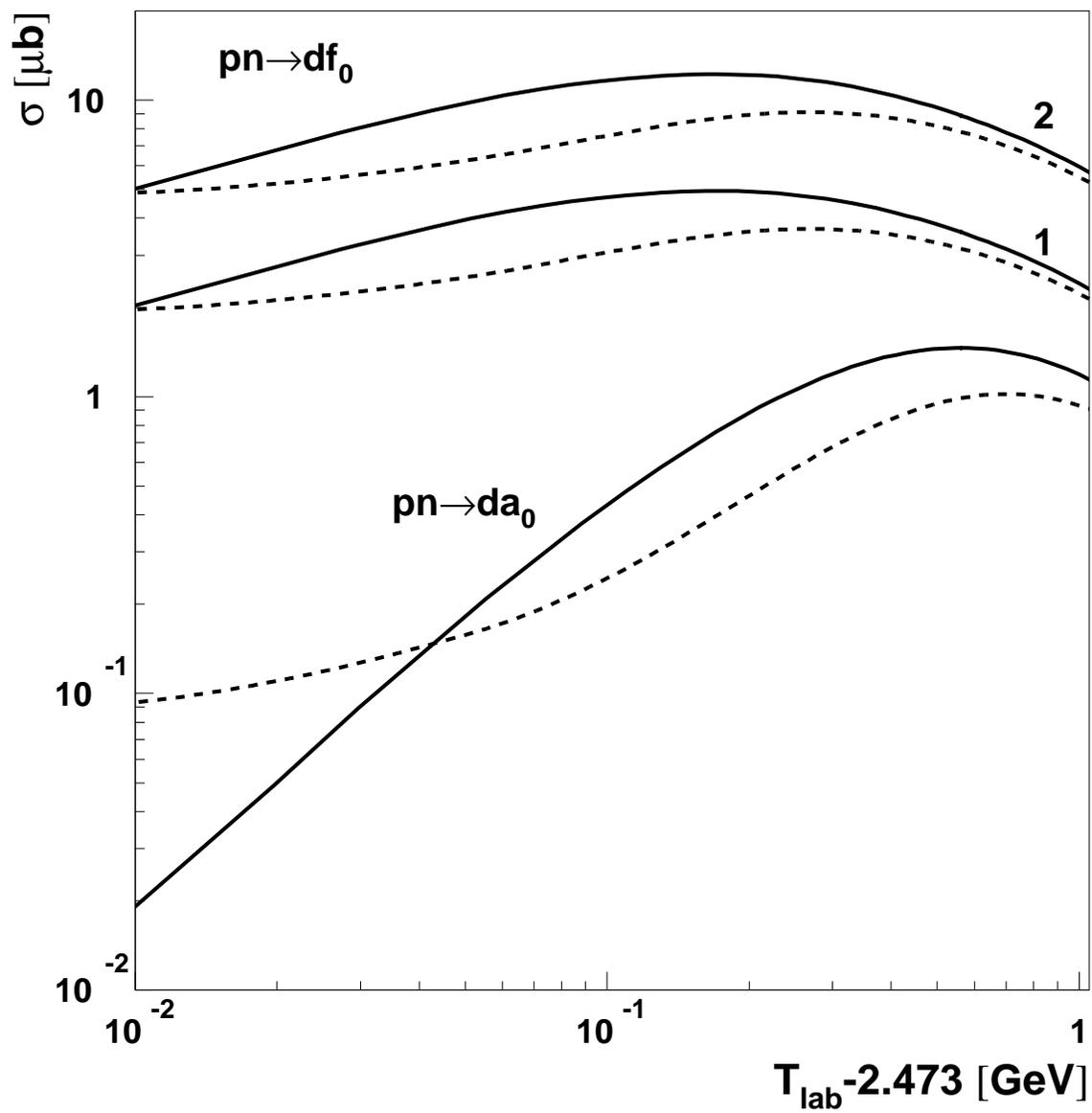,width=15cm}}
\vspace*{1cm}\caption{
Total cross sections for the reactions $pn \to da_0$ and $pn
\to df_0$ as a function of ($T_{\mathrm{lab}}-2.473$) GeV. The
solid and dashed curves are calculated using narrow and finite
resonance widths, respectively. The curves denoted by 1 and 2
correspond to the choices $R(f_0/a_0) $= 1.46 and 2.3, respectively.
}\label{fig:siga0f0}
\end{figure}

\begin{figure}[h]
\centerline{\psfig{file=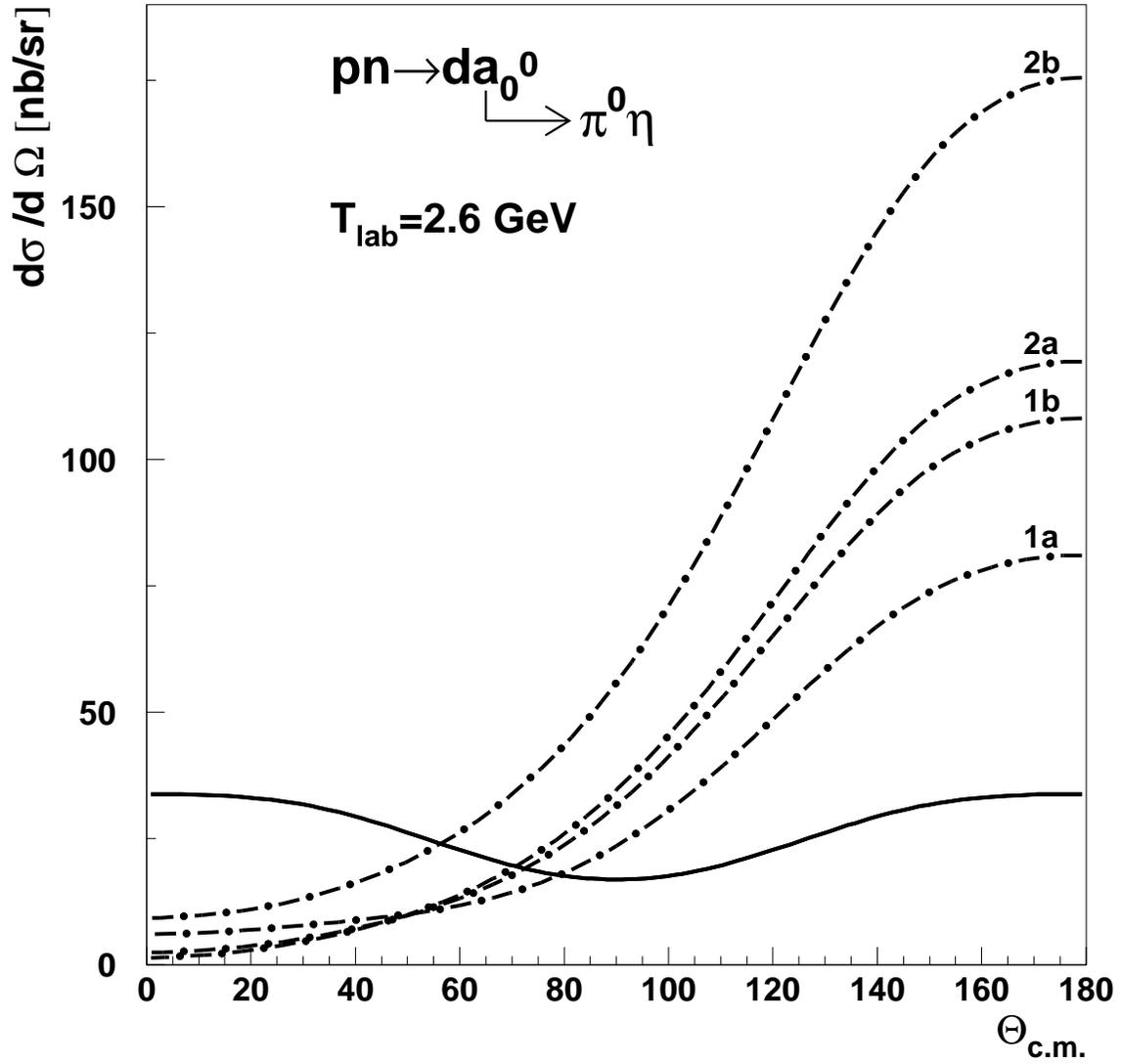,width=15cm}}
\vspace*{1cm}\caption{
Differential cross section of the reaction $pn \to d a_0^0$ at
$T_p=2.6$ GeV as a function
of $\Theta_{\mathrm{c.m.}}$. The solid curve corresponds to the
case of isospin conservation, i.e.\  $|\xi|^2=0$. The
dashed-dotted lines include the mixing effect with $|\xi|^2$ =
0.05 for the lower curves (1a and 2a) and $|\xi|^2 = 0.11$ for
the upper curves (1b and 2b). The lines 1a, 1b and 2a, 2b have
been calculated for $R(f_0/a_0) = 1.46$ and 2.3, respectively.
}\label{fig:dsa0f0}
\end{figure}

   The dash-dotted line in Fig.~\ref{fig:MM} shows our
estimations of possible background from nonresonant $\pi^0 \eta$
production in the reaction $pn \rightarrow d \pi^0 \eta$ at
$T_{{\rm lab}}=2.6$ GeV (see also  \cite{AnnRep2000}).
The background amplitude was described by the diagram shown
in Fig. \ref{fig:tsm} e), where $\eta$ and $\pi$ mesons are created
through the intermediate production of $\Delta(1232)$ (in the amplitude
$\pi N \rightarrow \pi N$) and $N(1535)$ (in the amplitude
$\pi N \rightarrow \eta N$). The total cross section of the nonresonant
$\pi \eta$ production due to this mechanism was found to be
$\sigma_{{\rm bg}} \simeq $ 0.8  $\mu$b for a cut-off in the
one-pion exchange $\Lambda_{\pi}= 1$ GeV.

The background is charge-symmetric and cancels in the difference of the
cross sections $ \sigma(\Theta) - \sigma(\pi - \Theta) $.
Therefore, the complete separation of the background is not
crucial for a test of isospin violation due to the $a_0$--$f_0$ mixing.
There will be also some contribution from
$\pi$-$\eta $ mixing as discussed in
\cite{Tippens,Magiera}. According to the results of \cite{Tippens}
this mechanism yields a charge-symmetry breaking in the
$\eta NN$ system of about 6\%:
$$R=d\sigma (\pi^+d \rightarrow pp\eta)/\sigma (\pi^-d \rightarrow
nn\eta)=~0.938 \pm 0.009.$$
A similar isospin violation due to
$\pi$-$\eta $ mixing can also be expected in our case.

The best strategy to search for isospin violation is a measurement of
the forward--backward asymmetry for different intervals of $M_{\eta
\pi^0}$.  As it follows from Fig. \ref{fig:MM} we have $\sigma_{a_0}
(\sigma_{{\rm bg}})= 0.3(0.4),~0.27(0.29)$ and 0.19(0.15) $\mu$b for
$M_{\eta \pi^0} \geq 0.85,~0.9$ and 0.95 GeV, respectively.  For
$M_{\eta \pi^0} \leq$ 0.7 GeV the resonance contribution is rather
small and the charge-symmetry breaking will be mainly related to
$\pi$-$\eta $ mixing and, therefore, will be small.  On the other hand,
in the interval $M \geq$ 0.95 GeV the background does not exceed the
resonance contribution and we expect a comparatively large isospin
breaking due to $a_0$-$f_0$ mixing.

\subsection{Reaction $pn \rightarrow d f_0 \rightarrow d \pi \pi $}

\begin{figure}[h]
\centerline{\psfig{file=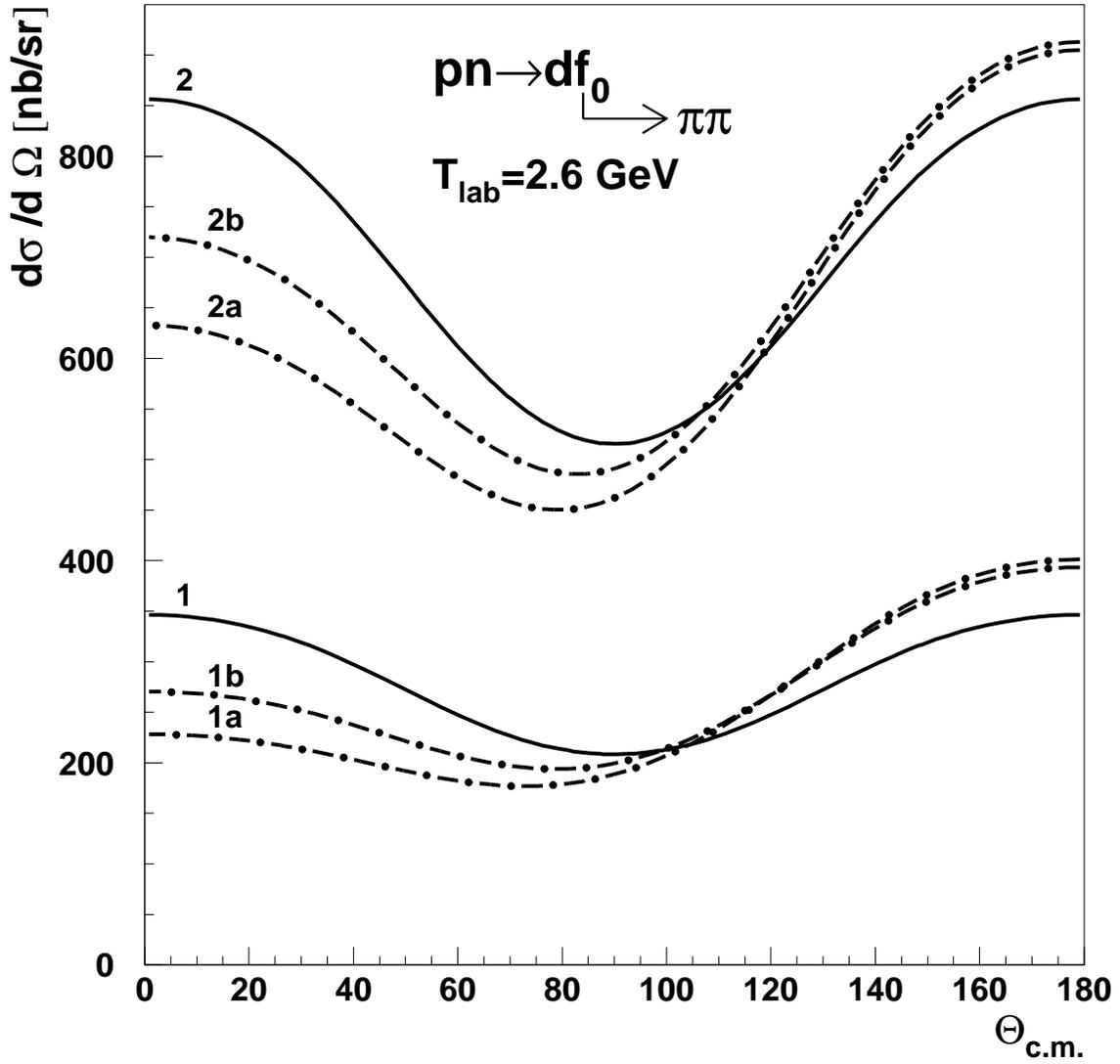,width=15cm}}
\vspace*{1cm}\caption{
Differential cross section of the reaction $pn \to d f_0$ at
$T_p=2.6$ GeV as a function of $\Theta_{\mathrm{c.m.}}$. The
notation of the curves is the same as in Fig. \ref{fig:dsa0f0}.
}\label{fig:dsf0a0}
\end{figure}

   The isospin-violation effects can also be measured in the reaction
\begin{equation}
pn \to d f_0 \to d \pi^+ \pi^- , \label{f0pipi}
\end{equation}
where, due to mixing, the $f_0$ may also be produced via the $a_0$.
The corresponding differential cross section is shown in
Fig.~\ref{fig:dsf0a0}. The differential cross section for $f_0$
production is expected to be essentially larger than for $a_0$
production, but the isospin violation effect turns out to be smaller
than in the $\pi \eta$-production channel. Nevertheless, the isospin-
violation parameter $A$ is expected to be about 10$-$30\% and can
be detected experimentally.

\subsection{Reactions $pd \rightarrow \mathrm{^3H}\, a_0^+$
             and $pd \rightarrow \mathrm{^3He}\, a_0^0$}

   We continue with $pd$ reactions and compare the final states
$\mathrm{^3H}\, a_0^+$ (c) and $\mathrm{^3He}\, a_0^0$ (d). Near
threshold the amplitudes of these reactions can be written as

\begin{eqnarray}
&&T(pd \rightarrow \mathrm{^3H}~ a_0^+)
=\sqrt{2} D_a \,{\bf S}_A \cdot {\bf e} ,
\end{eqnarray}
\begin{equation}
T(pd \rightarrow \mathrm{^3He}~ a_0^0)=(D_a+ \xi D_f){\bf S}_A \cdot {\bf e},
\end{equation}
with ${\bf S}_A=\phi_A^T \sigma_2 \
\mbox{\boldmath$\sigma$}\phi_N$. $D_a$ and $D_f$ are the scalar
$S$-wave amplitudes describing the $a_0$ and $f_0$ production in case
of $\xi$=0. The ratio of the differential cross sections for
reactions (d) and (c) is then given by
\begin{equation}
R_{dc}=\frac{|D_a+ \xi D_f|^2}{2|D_a|^2} = \frac12+
\frac{2 \mathrm{Re}(D_a^* \xi D_f)+|\xi D_f|^2}{|D_a|^2}.
\label{R_dc}
\end{equation}
The magnitude of the ratio $R_{dc}$ now depends on the relative
value of the amplitudes $D_a$ and $D_f$. If they are comparable
($|D_a| \sim |D_f|$) or $|D_f|^2 \gg |D_a|^2$ the deviation of
$R_{dc}$ from 0.5 (which corresponds to isospin conservation)
might be 100\% or more. Only in the case $|D_f|^2 \ll |D_a|^2$ the
difference of $|R_{dc}|^2$ from 0.5 will be small. However, this
seems to be very unlikely.

Using the two-step model for the reactions $pd \rightarrow
\mathrm{^3He}~ a_0^0$ and $pd \rightarrow \mathrm{^3He}~ f_0$,
involving the subprocesses $pp \rightarrow d \pi^+$ and $\pi^+ n
\rightarrow p~ a_0/f_0$ (cf.  \cite{Faldt,Uzikov}), we find
\begin{equation}
\frac{\sigma(pd \rightarrow \mathrm{^3He}~a_0^0)}{\sigma(pd
\rightarrow\mathrm{^3He}~f_0)} \simeq
\frac{\sigma(\pi^+ n\rightarrow p~a_0^0)}{\sigma(\pi^+ n\rightarrow p~f_0)}.
\end{equation}
According to the calculations in \cite{Grishina3} we expect
$\sigma(\pi^+ n \rightarrow pa_0^0)=\sigma(\pi^- p \rightarrow
na_0^0) \simeq 0.5$-$1$~mb at 1.75--2 GeV$/c$. A similar value for
$\sigma(\pi^- p \rightarrow n f_0)$ can be found using the results
from \cite{Brat99}. According to the latter study $\sigma(\pi^-
p\rightarrow nf_0 \rightarrow nK^+K^-) \simeq 6-8\ \mu$b at 1.75--2
GeV$/c$ and ${\rm Br}(f_0 \to K^+ K^-)\simeq 1\%$, which implies that
$\sigma(\pi^- p\rightarrow nf_0) \simeq$ 0.6--0.8~mb. Thus we
expect that near threshold $|D_a| \sim |D_f|$ . This would imply
that the effect of isospin violation in the ratio $R_{dc}$ can
become quite large.

Recently, the cross section of the reaction $pd\rightarrow
\mathrm{^3He}~K^+K^-$ has been measured by the MOMO collaboration at
COSY (J\"ulich) \cite{MOMO}. It was found $\sigma =9.6 \pm 1.0$ nb and $17.5 \pm 1.8$ nb
for $Q= 40$ and 56 MeV, respectively.  The authors note that the
invariant $K^+K^-$ mass distributions in those data contain a broad
peak which follows phase space.  However, as it was shown in
\cite{Brat01} the form of the invariant mass spectrum, which follows
phase space, can not be distinguished from the $a_0$ resonance
contribution at such small Q.  Therefore, the events from the broad
peak in \cite{MOMO} can also be related to the $a_0$ and/or $f_0$.
Moreover, due to the phase-space behavior near the threshold one would
expect a dominance of two-body reactions. Thus the real cross
section of the reaction $pd \rightarrow \mathrm{^3He}~a_0^0 \rightarrow
\mathrm{^3He}~\pi^0 \eta$ is expected to be not essentially smaller
than its upper limit of about 40$-$70 nb at $Q$ = 40--60 MeV which
follows from the MOMO data \cite{MOMO}.

\subsection{Reaction $dd\rightarrow \mathrm{^4He}\, a_0^0$}

   The direct production of the $a_0$ in the reaction $dd \rightarrow
\mathrm{^4He}\, a_0^0$ is forbidden. It thus can only be observed due
to the $f_0$--$a_0$ mixing:
\begin{equation}
\frac{\sigma(dd\rightarrow \mathrm{^4He}\, a_0^0)}
{\sigma(dd\rightarrow \mathrm{^4He}\, f_0)}= |\xi|^2.
\end{equation}
Therefore it will be very interesting to study the reaction
\begin{equation}
dd \rightarrow \mathrm{^4He}\, (\pi^0~ \eta) \label{dd}
\end{equation}
near the $f_0$-production threshold. Any signal of the reaction
(\ref{dd}) then will be related to isospin breaking. It is
expected to be much more pronounced near the $f_0$ threshold as
compared to the region below this threshold.

In summarizing this Section, we have discussed the effects of isospin
violation in the reactions $pN \rightarrow da_0$, $pn \rightarrow d f_0$
$pd \rightarrow \mathrm{^3He/^3H}\, a_0$ and $ dd \rightarrow \mathrm{^4He}\, a_0$
which can be generated by $f_0$--$a_0$ mixing. It has been demonstrated
that for a mixing intensity of about ($8\pm3$)\%, the isospin violation
in the ratio of the differential cross sections of the reactions $pp
\to da_0^+ \to d \pi^+ \eta$ and $pn \to da_0^0 \to d \pi^0 \eta$ as
well as in the forward--backward asymmetry in the reaction $pn \to
da_0^0 \to d \pi^0 \eta$ not far from threshold may be about 50--100\%.
Such large effects are caused by the interference of direct $a_0$
production and its production via the $f_0$ (the former amplitude is
suppressed close to threshold due to the $P$-wave amplitude whereas the
latter is large due to the $S$-wave mechanism).  A similar isospin
violation is expected in the ratio of the differential cross sections
of the reactions $pd \rightarrow \mathrm{^3H}\, a_0^+(\pi^+ \eta)$ and
$pd \rightarrow \mathrm{^3He}\, a_0^0(\pi^0\eta)$.  Finally, we have
also discussed the isospin violation effects in the reactions $pn \to
df_0(\pi^+\pi^-)$ and $ dd \rightarrow \mathrm{^4He}\, a_0$.  All
reactions together --- once studied experimentally --- are expected to
provide detailed information on the strength of the $f_0/a_0$ mixing.
Corresponding measurements are now in preparation for the ANKE
spectrometer at COSY (J\"ulich) \cite{proposal}.

\section*{Acknowledgements}
   The authors are grateful to J. Ritman for stimulating discussions and
useful suggestions and to V. Baru for providing the parametrization of the
FSI enhancement factor.
This work is supported by Deutsche Forschungsgemeinschaft
and by Russian Foundation for Basic Research.


\newpage
\begin{table*}[t]
\caption{\label{Tab1} Coefficients in Eq. (\ref{NNa0sum})
for different mechanisms of the $pp\to pp a_0^0$,
$pp \to pn a_0^+$, $pn\to pp a_0^-$ and  $pn \to pn a_0^0$ reactions}
\vspace*{5mm}
\begin{center}
\begin{tabular}{l c c c c }
\hline
Reaction $j$ (mechanism $\alpha$)
& $\xi^{\pi}_{j(\alpha)}[ab;cd]$ & $\xi^{\pi}_{j(\alpha)}[ab;dc]$ &
$\xi^{\pi}_{j(\alpha)}[ba;dc]$ &  $\xi^{\pi}_{j(\alpha)}[ba;cd]$
\\ \hline
$pp\to pp a_0^0 \ (t(\eta),t(f_1))$
& $+1/\sqrt{2}$ & $-1/\sqrt{2}$ & $+1/\sqrt{2}$ & $-1/\sqrt{2}$
\\
$\hphantom{pp\to pp a_0^0 }\ (s(N))$
& $+1/\sqrt{2}$ & $-1/\sqrt{2}$ & $+1/\sqrt{2}$ & $-1/\sqrt{2}$
\\
$\hphantom{pp\to pp a_0^0 }\ (u(N))$
& $+1/\sqrt{2}$ & $-1/\sqrt{2}$ & $+1/\sqrt{2}$ & $-1/\sqrt{2}$
\\
$\hphantom{pp\to pp a_0^0 }\ \mathrm{Regge}$
& $0$ & $0$ & $0$ & $0$
\\
\hline
$pp\to pn a_0^+ \ (t(\eta),t(f_1))$
& $-\sqrt{2}$ & $0$ & $0$ & $+\sqrt{2}$
\\
$\hphantom{pp\to pp a_0^0 }\ (s(N))$
& $0$ & $+\sqrt{2}$ & $-\sqrt{2}$ & $0$
\\
$\hphantom{pp\to pp a_0^0 }\ (u(N))$
& $+2\sqrt{2}$ & $-\sqrt{2}$ & $+\sqrt{2}$ & $-2\sqrt{2}$
\\
$\hphantom{pp\to pp a_0^0 }\ \mathrm{Regge}$
& $-1$ & $+1$ & $-1$ & $+1$
\\
\hline
$pn\to pp a_0^- \ (t(\eta),t(f_1))$
& $+1$ & $-1$ & $0$ & $0$
\\
$\hphantom{pp\to pp a_0^0 }\ (s(N))$
& $-2$ & $+2$ & $-1$ & $+1$
\\
$\hphantom{pp\to pp a_0^0 }\ (u(N))$
& $0$ & $0$ & $+1$ & $-1$
\\
$\hphantom{pp\to pp a_0^0 }\ \mathrm{Regge}$
& $+1/\sqrt{2}$ & $-1/\sqrt{2}$ & $-1/\sqrt{2}$ &
$+1/\sqrt{2}$
\\
\hline
$pn\to pn a_0^0 \ (t(\eta),t(f_1))$
& $-1$ & $0$ & $+1$ & $0$
\\
$\hphantom{pp\to pp a_0^0 }\ (s(N))$
& $-1$ & $-2$ & $+1$ & $+2$
\\
$\hphantom{pp\to pp a_0^0 }\ (u(N))$
& $-1$ & $+2$ & $+1$ & $-2$
\\
$\hphantom{pp\to pp a_0^0 }\ \mathrm{Regge}$
& $0$ & $+\sqrt{2}$ & $0$ & $-\sqrt{2}$
\\
\hline
\end{tabular}
\end{center}
\end{table*}


\clearpage

\clearpage

\clearpage

\clearpage

\clearpage

\clearpage

\clearpage

\clearpage

\clearpage

\clearpage

\clearpage

\clearpage

\clearpage

\clearpage

\clearpage

\clearpage

\end{document}